%% file: main.tex
\documentclass[useAMS,usenatbib]{mn2e}
\usepackage[a4paper,margin=2cm]{geometry}
\usepackage[pdftex]{graphicx}
\pdfoutput=1
\usepackage{color}
\usepackage{aas_macros}
\usepackage{lastpage}
\usepackage{comment}
\usepackage{amsmath}
\usepackage{type1cm}
\usepackage{eso-pic}
\usepackage{url}

\input{commands}

\title[DRAGONS I: Dynamical lives of high-z galaxies]{Dark-ages~Reionization~$\&$~Galaxy~Formation~Simulation~I:\\The dynamical lives of high redshift galaxies}

\author[Poole et al.]{\parbox[t]{\textwidth}{
    Gregory B.\ Poole$^1$\footnotemark,
    Paul W. Angel$^1$,
    Simon J. Mutch$^1$,
    Chris Power$^2$, \\
    Alan R. Duffy$^{1,3}$,
    Paul M. Geil$^1$,
    Andrei Mesinger$^4$,
    Stuart B. Wyithe$^1$
}\\ \\
  $^1$ School of Physics, University of Melbourne, Parksville, VIC 3010, Australia \\
  $^2$ ICRAR, University of Western Australia, 35 Stirling Highway, Crawley, Western Australia 6009, Australia \\
  $^3$ Centre for Astrophysics \& Supercomputing, Swinburne University of Technology, P.O. Box 218, Hawthorn, VIC 3122, Australia \\
  $^4$ Scuola Normale Superiore, Piazza dei Cavalieri 7, I-56126 Pisa, Italy 
}
\date{draft version \today}
\pagerange{\pageref{firstpage}--\pageref{lastpage}} \pubyear{2012}

\begin{document}

\label{firstpage}

\maketitle

\begin{abstract}
\input{abstract}
\end{abstract}


\section{Introduction}\label{sec-intro} 
\renewcommand{\thefootnote}{\fnsymbol{footnote}}
\setcounter{footnote}{1}
\footnotetext{E-mail: gpoole@unimelb.edu.au}

\input{introduction}\label{sec-introduction}

\section{Simulations}\label{sec-simulations}
\input{simulations}

\section{Analysis}\label{sec-analysis}
\input{analysis}

\section{Summary and conclusions}\label{sec-summary}
\input{summary}

\section*{Acknowledgements}
\input{thanks}

\bibliographystyle{mn2e}
\bibliography{main}

\appendix
\section{Mass function fitting}\label{sec-appendix}
\input{appendix}

\end{document}

%% file: commands.tex
\def\lesssim{\mathrel{\hbox{\rlap{\hbox{\lower4pt\hbox{$\sim$}}}\hbox{$<$}}}}
\def\gtrsim{\mathrel{\hbox{\rlap{\hbox{\lower4pt\hbox{$\sim$}}}\hbox{$>$}}}}
\newcommand{\ltaraw}{$\; \buildrel < \over \sim \;$}
\newcommand{\lta}{\lower.5ex\hbox{\ltaraw}}
\newcommand{\gtaraw}{$\; \buildrel > \over \sim \;$}
\newcommand{\gta}{\lower.5ex\hbox{\gtaraw}}
\def\lsim{\mathrel{\rlap{\lower4pt\hbox{\hskip1pt$\sim$}}
    \raise1pt\hbox{$<$}}}                

\newcommand{\quotes}[1]{``#1''}

\newcommand{\ie}{{\rm i.e.}}
\newcommand{\eg}{{\rm e.g.}}

\newcommand{\etal}{{\rm et~al.}}

\newcommand{\HII}{\textsc{Hii}}

\def\apjl{ApJ}%
\def\apjs{ApJS}%
\def\ao{Appl.~Opt.}%
\def\aap{A\&A}%
\newcommand{\TWOLPTIC}{\textsc{2LPTic}}
\newcommand{\GADGETtwo}{\textsc{GADGET-2}}
\newcommand{\Subfind}{\textsc{Subfind}}
\newcommand{\ROCKSTAR}{\textsc{ROCKSTAR}}
\newcommand{\Meraxes}{\textsc{Meraxes}}

\newcommand{\Tiamat}{\textit{Tiamat}}
\newcommand{\Smaug}{\textit{Smaug}}
\newcommand{\TinyTiamatW}{\textit{Tiny Tiamat-W07}}
\newcommand{\TinyTiamat}{\textit{Tiny Tiamat}}
\newcommand{\MediTiamat}{\textit{Medi Tiamat}}
\newcommand{\GiggleZ}{\textit{GiggleZ}}

\newcommand{\GiggleZHR}{\textit{GiggleZ-HR}}

\newcommand{\Mvir}{{${\rm M}_{\rm vir}$}}

\newcommand{\tauthreetoone}{{$\tau_{\rm 3:1}$}}
\newcommand{\tautentoone}{{$\tau_{\rm 10:1}$}}
\newcommand{\tauform}{{$\tau_{\rm form}$}}
\newcommand{\taumerge}{{$\tau_{\rm merge}$}}

\newcommand{\xoff}{{$x_{\rm off}$}}
\newcommand{\fsub}{{$f_{\rm sub}$}}
\newcommand{\Vir}{{$\varphi$}}

\newcommand{\Mpc}{\,\mbox{Mpc}}
\newcommand{\Mpchunit}{$h^{-1} {\rm Mpc}$}
\newcommand{\kpchunit}{$h^{-1} {\rm kpc}$}

\newcommand{\Myr}{\,\mbox{Myr}}

\newcommand{\yr}{\,\mbox{yr}}
\newcommand{\kms}{\, \mbox{km}\,\mbox{s$^{-1}$}}

\newcommand{\Msol}{${{\rm M}_{\sun}}$}
\newcommand{\Msolhunit}{${{h^{-1}}{\rm M}_{\sun}}$}

\newcommand{\Myrs}{\,\mbox{Myrs}}


%% file: abstract.tex
We present the Dark-ages Reionization and Galaxy-formation Observables from Numerical Simulations (DRAGONS) program and \Tiamat, the collisionless N-body simulation program upon which DRAGONS is built.  The primary trait distinguishing \Tiamat\ from other large simulation programs is its density of outputs at high redshift (100 from $z{=}35$ to $z{=}5$; roughly one every 10\Myrs) enabling the construction of very accurate merger trees at an epoch when galaxy formation is rapid and mergers extremely frequent. We find that the friends-of-friends halo mass function agrees well with the prediction of \citet{Watson:2013p2555} at high masses, but deviates at low masses, perhaps due to our use of a different halo finder or perhaps indicating a break from \quotes{universal} behaviour.  We then analyse the dynamical evolution of galaxies during the Epoch of Reionization finding that only a small fraction (${\sim}20$\%) of galactic halos are relaxed.  We illustrate this using standard relaxation metrics to establish two dynamical recovery time-scales: \textit{i}) halos need ${\sim}1.5$ dynamical times following formation, and \textit{ii}) ${\sim}2$ dynamical times following a major (3:1) or minor (10:1) merger to be relaxed.  This is remarkably consistent across a wide mass range.  Lastly, we use a phase-space halo finder to illustrate that major mergers drive long-lived massive phase-space structures which take many dynamical times to dissipate.  This can yield significant differences in the inferred mass build-up of galactic halos and we suggest that care must be taken to ensure a physically meaningful match between the galaxy-formation physics of semi-analytic models and the halo finders supplying their input.

%% file: introduction.tex
Following cosmological recombination the baryonic gas filling the Universe became predominantly neutral. The fact that this gas is known to be mostly ionized today \citep{Gunn:1965p2569} implies that the intergalactic medium (IGM) under went a significant {\em reionization} event at some early point in its history. This fact is responsible for some of the major questions in extragalactic astronomy including: when did this process occur and what were the responsible ionizing sources? Recent observations have begun to provide preliminary answers to this question \citep[\eg][]{Fan:2006p2573,Ouchi:2010p2575,PlanckCollaboration:2015p2562}.  Soon measurements of highly redshifted 21-cm radio emission \citep{Furlanetto:2006p2567,Morales:2010p2568} will open an important new observational window for study of the first galaxies, providing the first direct probe of the neutral hydrogen content in the early Universe.

The development of theoretical models that self-consistently include the physics of galaxy formation and intergalactic hydrogen will play a key role in understanding the nature of the first galaxies and in interpreting these observations.  This paper is the first in a series describing the Dark-ages Reionization and Galaxy-formation Observables from Numerical Simulations (DRAGONS\footnote{\url{http://dragons.ph.unimelb.edu.au/}}) project which aims to integrate detailed semi-analytic models constructed specifically to study galaxy formation at high redshift, with semi-numerical models of the galaxy--IGM interaction \citep{Zahn:2007p2570,Mesinger:2007p2571,Geil:2008p2572}.  The galaxy-formation modelling for DRAGONS is implemented using a set of large N-body simulations which we refer to as the \Tiamat\ simulation suite.  \Tiamat\ provides a framework within which to implement a semi-analytic model for reionization and to study the formation histories, structure and properties of the dark matter halos that dictate the formation sites and assembly histories of the first galaxies. 

Over the past decade the requirements for simulations aiming to address the structure of reionization and galaxy formation in the Epoch of Reionization (EoR) have been studied extensively.  The consensus from previous N-body studies \citep[\eg][]{Iliev:2007p2581,Zahn:2007p2570,McQuinn:2007p2582,Shin:2008p2583,Lee:2008p2584,Croft:2008p2585} and analytic models \citep[\eg][]{Furlanetto:2004p2578,Wyithe:2007p2579,Barkana:2009p2580} is that large-scale over-dense regions near bright sources ionize first with clustered neighbouring sources contributing to increase the size of ionised regions. Simulations on the scale of 100\Mpc\ are found to be large enough to correctly capture the structure and duration of reionization, although volumes up to 500$^3$\Mpc\ are required to capture all large scale power due to clustering of \HII\ regions \citep{Iliev:2014p2577}.

The challenge is to model galaxy formation in volumes of this size with sufficient resolution.  In the cold neutral IGM prior to reionization, molecular cooling may proceed within minihalos with masses ${\sim}10^6$\Msol.  However, the processes principally responsible for regulating galaxy formation are expected to be active in halos with virial temperatures greater than $T_{\rm min}{\sim}10^4$K, above which atomic hydrogen cooling becomes efficient.  On the other hand, the growth of \HII\ regions during reionization is also expected to be influenced by radiative feedback due to suppression of galaxy formation below the cosmological Jeans mass within a heated IGM \citep[\eg][]{Dijkstra:2004p2576}. Together these constraints indicate that sufficient resolution is required to identify halo masses down to ${\sim}5{\times}10^7 {\rm M_\odot}$ solar masses within a volume of ${\sim}100$\Mpc. 

In addition to this dynamic range of scales, for the DRAGONS program we aim to accurately resolve the relevant time-scales of high-redshift galaxy formation putting an additional constraint on the cadence with which simulation outputs must be generated.  At $z{\sim}6$ the dynamical time of a galactic disc falls below the lifetime of the least massive Type-II supernova progenitor (${\sim}4{\times}10^7$\yr).  As a result, snapshots with a cadence of ${\sim}10^7$ years are required to follow galaxy formation correctly during the EoR with a semi-analytic model.  This interval is an order of magnitude shorter than needed to describe galaxy formation at redshifts $z{\sim}0$.

In this paper we present the \Tiamat\ suite of collisionless N-body simulations which we have run to satisfy these requirements and upon which the DRAGONS program will be constructed.  Given its critical importance as the foundation of the program, we take this opportunity to present the methodology of constructing this set of simulations and to characterise the populations of galactic halos obtained.  In particular, we shall carefully examine the dynamical evolution of galactic halos during the reionization era.  We seek this understanding because of its potential impact on the structure of galactic halos, which is of fundamental importance to the physics of galaxy formation at any epoch, including the EoR.  In particular, halo concentrations and angular momenta are generally believed to dictate the size and surface density of the disc-like structures in which the majority of star formation occurs.  Dynamical disturbances can additionally drive starbursts or affect the stability of these disc-like structures, strongly affecting star formation and forcing morphological transformations which contribute to the assembly of galactic spheroids.  This can in turn affect observable galaxy sizes or alter UV fluxes and escape fractions with important effects on the reionization history of the Universe.

At low redshift, it has been shown that a halo's dynamical state can systematically affect the structure and gravitational potential of galactic halos \citep[\eg][]{Thomas:1998p2559,Neto:2007p2556,Power:2012p2560,Ludlow:2014p2558}.  These studies have collectively established a set of criteria (which we refer to henceforth as \quotes{standard} relaxation criteria) capable (at low redshifts at least) of separating halos with disturbed structure from those with relaxed structure.  These standard criteria consist of cuts on three metrics for each halo: the separation of its dense centre from its centre of mass (\xoff), its \quotes{pseudo-virial ratio} constructed from its velocity dispersion and gravitational binding energy (\Vir), and its substructure fraction (\fsub).  When low-redshift halos are separated into relaxed and unrelaxed samples in this way, substantial effects on halo concentration and (to a lesser extent) spin have been demonstrated.  This is of particular importance to studies which aim to understand the processes which establish the \quotes{universal} density profiles of halos extracted from collisionless N-body simulations \citep{Navarro:1997p2543}.

While there is broad agreement at low redshift as to the dependance of halo structure on mass, redshift and dynamical state, recent studies which have attempted to push this understanding to the EoR \citep[\eg][]{Prada:2012p2538,Diemer:2015p2657,Dutton:2014p2658,Hellwing:2015p2659} have found less consensus.  At high redshifts where simulations predict that merger rates are very high, halos significantly less concentrated, and merger orbital properties quite different, the influence of dynamical state on halo structure may differ from local trends.  It is unclear to what degree dynamical disturbance may play a role in the differences in high-redshift halo structure reported in the literature since these studies have not been consistent in their treatment of this issue.

Unfortunately, the details of how the standard relaxation metrics evolve following dynamical disturbances has not been properly explored at any redshift, nor has their efficacy at separating relaxed systems from unrelaxed systems been demonstrated at high redshift.  Before presenting a detailed analysis of halo structure at high redshift to understand the discrepancies present in the literature, we aim first to address both of these issues here.  Due to the historical focus on low redshifts by galaxy-formation models, few large simulation programs possess sufficient temporal resolution to perform a thorough dynamical analyses at high redshift.  Given its fine snapshot temporal resolution, \Tiamat\ represents a unique resource for exploring these issues across the full range of masses most relevant to galaxy formation in the early Universe.  We will find that the standard relaxation criteria are effective at identifying systems that are recovering from their formation or from recent significant mergers.  With this methodology properly validated at high redshift, we will subsequently perform a detailed analysis of the structure of both relaxed and unrelaxed high redshift dark matter halos -- including spin parameter and concentration-mass relations -- in a companion paper (Angel et al.~2015; PAPER-II).

The \Tiamat\ N-body simulation hosts a semi-analytic model of galaxy formation named \Meraxes, which has been integrated within a semi-numerical model for ionization structure. In subsequent papers we will present this model (Mutch et al.~2015; Paper-III) and use it in a range of studies including high redshift galaxy luminosity functions (Liu et al.~2015, PAPER-IV) and the ionization structure of the intergalactic medium (Geil et al~2015, PAPER-V).  Complementary high resolution hydrodynamics simulations called \Smaug\ \citep[already presented in][]{Duffy:2014p2561} will characterise the basic scaling relationships of early galaxy formation. There will then be a detailed comparison of \Meraxes\ to the results of \Smaug\ with suggested constraints of the semi-analytic model based on hydrodynamics (Qin \etal, in prep).
 
The outline of this paper is as follows. In Section \ref{sec-simulations} we introduce the construction of the \Tiamat\ simulations including our approaches to halo finding and merger tree construction.  In Section \ref{sec-analysis} we analyse these data products to identify how galactic halos relax at high redshift, estimate their relaxed fraction and present some preliminary findings on their phase-space structure.  Finally, in Section \ref{sec-summary} we summarise our study and present our conclusions. Our choice of fiducial cosmology throughout will be a standard spatially-flat Planck $\Lambda$CDM cosmology based on 2015 data \citep[][]{PlanckCollaboration:2015p2562} ($h$, $\Omega_{\rm{m}}$, $\Omega_{\rm{b}}$, $\Omega_\Lambda$, $\sigma_8$, $n_{\rm{s}}$)${=}$($0.678$, $0.308$, $0.0484$, $0.692$, $0.815$, $0.968$) although we will make isolated use of two simulations run with standard spatially-flat WMAP-5 \citep{Komatsu:2009} ($h$, $\Omega_{\rm{m}}$, $\Omega_{\rm{b}}$, $\Omega_\Lambda$, $\sigma_8$, $n_{\rm{s}}$)${=}$($0.727$, $0.273$, $0.0456$, $0.705$, $0.812$, $0.96$) and WMAP-7 \citep{Komatsu:2011p1737} ($h$, $\Omega_{\rm{m}}$, $\Omega_{\rm{b}}$, $\Omega_\Lambda$, $\sigma_8$, $n_{\rm{s}}$)${=}$($0.702$, $0.275$, $0.0458$, $0.725$, $0.816$, $0.96$) $\Lambda$CDM cosmologies.

%% file: simulations.tex
\begin{table*}
\begin{minipage}{170mm}
\begin{center}
\begin{tabular}{|l|ccccc|cc}
\hline
Simulation  & $N_{\rm{p}}$  & L [Mpc/$h$] & $m_{\rm{p}}$ [$\rm{M}_{\odot}$/$h$] & $\epsilon$ [kpc/$h$] & $\eta$ & Cosmology & Halo Finding\\
\hline
Tiamat		   & $2160^3$ & 67.8 & $2.64 {\times} 10^6$ & 0.63	& 0.025 & Planck-2015	& \Subfind\\
Medi Tiamat	   & $1080^3$ & 22.6 & $7.83 {\times} 10^5$ & 0.42	& 0.025 & Planck-2015	& \Subfind\\
Tiny Tiamat	   & $1080^3$ & 10.0 & $6.79 {\times} 10^4$ & 0.19	& 0.025 & Planck-2015	& \Subfind\\ 
Tiny Tiamat-W07 & $1024^3$ & 10.0 & $7.11 {\times} 10^4$ & 0.20	& 0.010 & WMAP-07		& \Subfind\ \& \ROCKSTAR\\ 
\hline
\end{tabular}
\caption{Box sizes ($L$), particle counts ($N_p$), particle mass ($m_p$), gravitational softening lengths ($\epsilon$) and integration accuracy parameters ($\eta$) for the \Tiamat\ simulations as well as the cosmology and halo finding codes used for each. \label{table-simulation_parameters}}
\end{center}
\end{minipage}
\end{table*}

In this section we present our approach to assembling the \Tiamat\ suite of collisionless N-body simulations.  We describe the means by which initial conditions were generated, the simulation code used and its parameters and the means by which bound structures (halos) were extracted and assembled into merger trees for the analysis presented in subsequent sections.  A summary of the most essential metrics of our simulations is presented in Table \ref{table-simulation_parameters}.

\subsection{Simulation runs}\label{sec-sim_runs}

The simulation data products upon which the DRAGONS program is being built come primarily from \Tiamat; its flagship collisionless N-body simulation.  \Tiamat\ has been designed to provide sufficient mass resolution to accurately capture the low-mass galaxy population driving the reionization of the Universe at high redshifts and to do so over a sufficiently large volume to capture the evolving structure of the reionization field, right to the epoch of bubble overlap.  It does so with sufficiently high snapshot temporal resolution to fully capture the rapid evolution of both.  Specifically, \Tiamat\ consists of a periodic box, 100 \Mpc\ (comoving) on a side sampled with 2160$^3$ particles with 100 snapshots of particle data recorded at intervals equally spaced in cosmic time from $z{=}35$ to $z{=}5$ (\ie\ every 11.1 \Myr).  To facilitate several smaller-volume but higher resolution studies we have run a series of companion simulations -- each consisting of ${\sim}10^9$ particles -- named \TinyTiamat\ (10 \Mpchunit\ box) and \MediTiamat\ (22.6 \Mpchunit\ box) using the same snapshot cadence strategy as \Tiamat.  We have also run a companion 10 \Mpchunit\ box named \TinyTiamatW\ with a WMAP-7 cosmology.

All aspects of the simulations were performed assuming standard $\Lambda$CDM cosmologies with the parameters given at the end of Section \ref{sec-introduction}.  Initial conditions were generated with the 2nd-order Lagrangian perturbation code \TWOLPTIC\ \citep[][with fixes to ensure correct behaviour when particle or displacement field grid cell counts exceed $2^{32}{-}1$]{Crocce:2006p2494} at $z{=}99$ ($z{=}127$ for \TinyTiamatW) using a particle load with a regular periodic grid structure and a displacement field computed on regular periodic grids of dimensions 2160$^3$ for \Tiamat\ and 1080$^3$ for all other cases.  The input power spectrum was generated using CAMB \citep{Lewis:2000} with the $\Lambda$CDM parameters appropriate to each.

We have run our simulations using \GADGETtwo\ \citep{Springel:2005b}, a Tree-Particle Mesh (TreePM) code well suited to large distributed memory systems, with the RAM conserving modifications listed in \citet{Poole:2015a}.  Gravitational softening was set to 0.02$\bar{d}$ (where $\bar{d}=L/\sqrt[3]{N_p}$ is the mean interparticle spacing of a simulation with $N_p$ particles in a cubic volume of side length $L$) in all cases \citep[as motivated by][]{Poole:2015a} and the integration accuracy parameter was set to $\eta{=}0.025$ in all cases except \TinyTiamatW\ where $\eta{=}0.01$.  For all but \TinyTiamatW, this is a slight relaxation of the choice argued for in \citet{Poole:2015a} but was deemed necessary to reduce the wallclock time for the calculation to required levels.

Lastly, at several places we will seek to compare our high-redshift findings from \Tiamat\ to the dynamical activity of similarly sized systems at low redshift.  To facilitate this comparison, we will use the \GiggleZHR\ simulation (\ie\ the highest resolution \GiggleZ\ control volume simulation, see \citealt{Poole:2015a} for details).  This simulation consists of 1080$^3$ particles in a 125 \Mpchunit\ box and was run with a WMAP-5 cosmology.  Halo finding and tree building was performed in the same way and with the same code versions used for \Tiamat.  For the purposes of the work we present here, this simulation should provide an adequate comparison between our highest mass high-redshift \Tiamat\ halos and halos of similar mass in the low redshift Universe. 

\subsection{Halo finding and merger tree construction}\label{sec-halo_finding}

We have preformed the majority of our halo finding using the widely utilised \Subfind\ code of \citet{Springel:2001}.  This code initially identifies collapsed regions of interest using a friends-of-friends (FoF) algorithm (for which we use the standard linking length criterion of 0.2$\bar{d}$) and subsequently identifies bound substructures within these FoF groups as locally overdense collections of particles, removing unbound particles through an unbinding procedure.  Halo centres are taken to be the position of a halo's most bound particle as identified by \Subfind\ unless otherwise stated.

For the \TinyTiamatW\ simulation we have additionally performed the halo finding exercise using the publicly available version of the 6D phase-space code \ROCKSTAR\ \citep{Behroozi:2013p2552}.  This code also performs an initial FoF search for structures, this time with a generous linking length of 0.26$\bar{d}$, and subsequently searches these objects for distinct structures in phase space.  Careful examination of the particle list output from \ROCKSTAR\ revealed numerous cases where particles are assigned to multiple halos or whole halos are duplicated (involving approximately 0.1\% of particles at $z{=}5$ in \TinyTiamatW).  In the course of rewriting the particle lists to match the standard \Subfind\ format required by our analysis codes (\ie, particle lists organised by friends-of-friends group, in order of subgroup size and sorted in radial order from the centre of the system), we have eliminated these cases by removing duplicated halos and allocating particles with multiple halo assignments to the halo with the nearest centre as determined using the shrinking sphere method of \citet{Power:2003p2553}.

We have constructed merger trees from these halo catalogs following the method to be presented in Poole \etal\ (2015; in prep.).  This approach carefully repairs artefacts introduced by imperfections in the halo finding process and identifies pathologies in the merger trees (\eg\ instances when tree branches are broken due to overlinking by the halo finder or as a result of halo fragmentations) through a process of forward and backward matching which scans both ways over multiple snapshots.  For our \Tiamat\ trees, we have used 16 snapshots for this process (\ie\ $\Delta t_{\rm scan}{\sim}160$\Myrs) representing more than a full dynamical time even at $z{=}5$, which we find sufficient for an accurate calculation.

\subsection{Halo Mass function}\label{sec-mass_sub_functions}

In addition to halo accretion histories and merger trees, the evolving abundance of galactic halos in \Tiamat\ will be a primary determinant of the galaxy populations and reionization histories we derive from DRAGONS modelling efforts.  Parameterisations of these mass functions are of great utility for a wide variety of semi-numerical extragalactic calculations and are an excellent way to facilitate comparisons to halo populations from other studies in the literature.  For these reasons, and given the excellent combination of volume and mass resolution covered by \Tiamat\ at high redshift, we present here the \Tiamat\ halo mass functions and a parameterised fit to them.

A great deal of effort has been invested in the literature on methods of accurately and robustly estimating the halo mass function across a wide range of masses, cosmologies and redshifts and we refer the interested reader to the studies of \citet{Press:1974p2491,Jenkins:2001p2646,Lukic:2007,Reed:2007p2649,Tinker:2008p2551,Knebe:2013p2647,Murray:2013p2648} and references therein for a detailed account of the subject.  Our emphasis here is merely to present the halo mass function in the most recent Planck cosmology across ranges in mass and redshift relevant to galaxy formation during the EoR.  Given the recent interest in the \quotes{universal} friends-of-friends halo mass function parameterisation presented by \citet[][see Appendix \ref{sec-appendix} for details]{Watson:2013p2555}, we focus here on a parameterisation of that form.  We have used a Monte-Carlo Markov Chain (MCMC) approach to perform this fit and present the details of our approach, the resulting best fit parameters and their covariance in Appendix A.

\begin{figure}
\includegraphics[width=80mm]{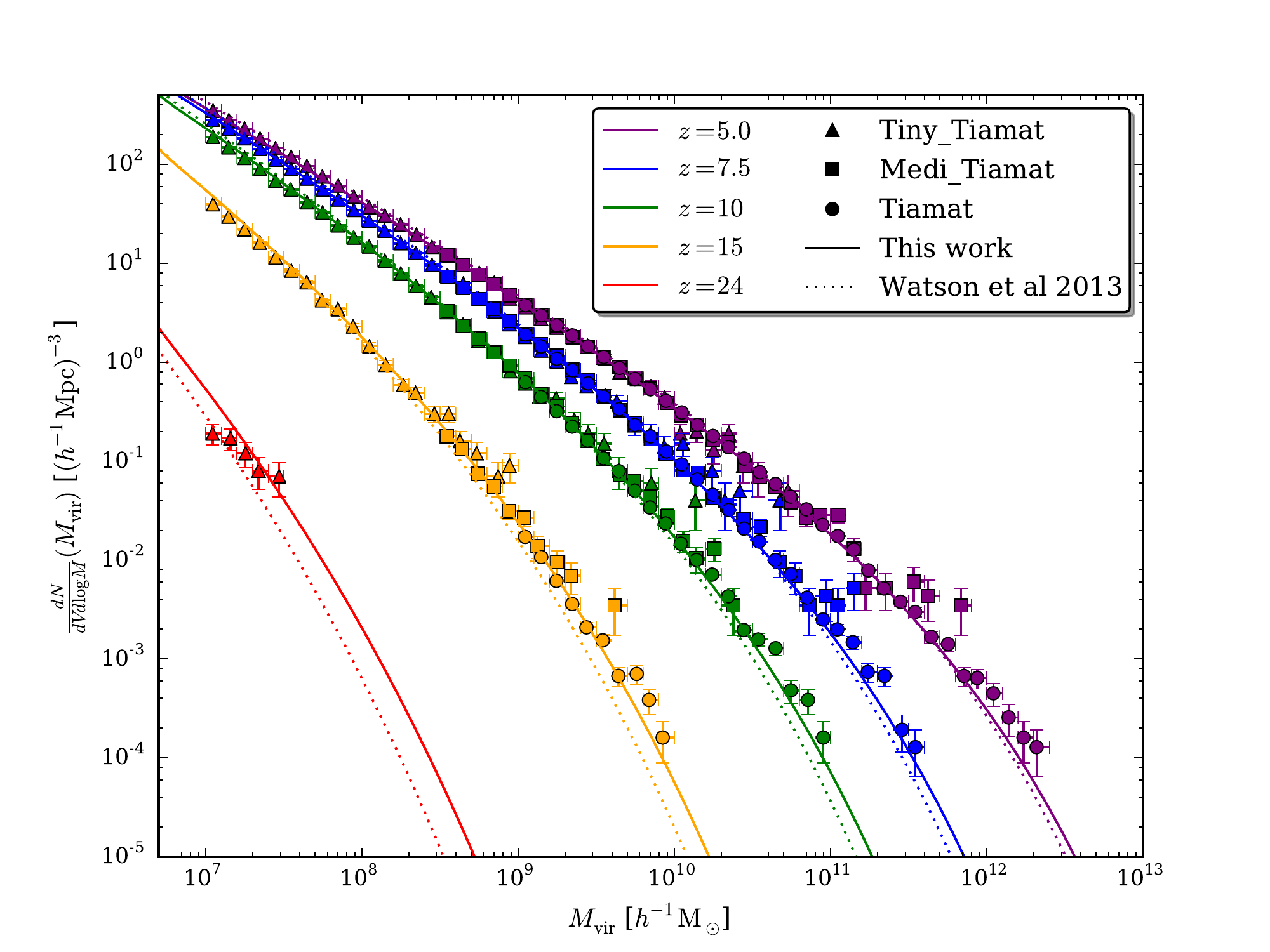}
\caption[Mass functions]{\Tiamat\ mass functions at redshifts $z{=}5$,$7.5$,$10$,$15$ and $24$ compared to the fitting formula of \citet{Watson:2013p2555}.  Dotted lines present results for the original fit presented in \citet{Watson:2013p2555} and solid lines present results of the the refitting performed in this study.\label{fig-mass_functions}} 
\end{figure}

In Figure \ref{fig-mass_functions} we present the FoF halo mass functions derived from the \Subfind\ catalogs extracted from the \TinyTiamat, \MediTiamat\ and \Tiamat\ simulations and compare these to the 4-parameter \quotes{universal} FoF mass function of \citet{Watson:2013p2555} as fit originally in that work (dotted lines) and as refitted in this work (solid lines).  Some small but significant differences are found between the original \citet{Watson:2013p2555} parameterisation and our re-parameterisation, particularly at the lowest masses and highest redshifts.  There are many possible sources for this including numerical reasons, such as those arising from systematic differences between our halo finder and that employed in the \citet{Watson:2013p2555} study, as well as more physical reasons such as systematic differences in FoF linking during the EoR when the most massive halos lie in highly filamentary regions where structures are more prone to overlinking by this algorithm.  All of our data for the \Tiamat\ fits is at high redshift while the \citet{Watson:2013p2555} fits are to data spanning a much wider range of redshifts, so the difference may also simply reflect the fact that the FoF halo mass function deviates from a universal form, at least at high redshifts.  Given the important contributions low-mass galaxies make to the ionizing photon budget during the EoR, such differences are important to note.

%% file: analysis.tex
In this section we examine the time-scales by which galactic halos at high redshift relax following formation and mergers.  We relate these time-scales to their dynamical age and to intervals between merger events and will find that only towards the end of the epoch of reionization do significant numbers of halos exist in relaxed states.  Lastly, we will also present some preliminary results about their phase-space structure that may be of consequence for the application of semi-analytic models at high redshift.

\subsection{Dynamical recovery time-scales}\label{sec-recovery_timescales}

It has long since been shown (and perhaps not surprising) that the structure of halos extracted from collisionless N-body simulations has a significant dependance on the dynamical state of the system \citep{Thomas:1998p2559,Neto:2007p2556,Power:2012p2560,Ludlow:2014p2558}.  At low redshifts at least, cuts on three metrics for quantifying the dynamical state of halos have found success at separating systems with disturbed structure from those with relaxed structure: the offset parameter (\xoff), given by the displacement of the densest centre of a halo from its centre of mass; the virial ratio (\Vir), given by $2K{/}\left|U\right|$, where $K$ is the kinetic energy and $U$ is the halo's gravitational binding energy \citep[see Section 5.1 of][and references therein, for a detailed description of virialisation]{Poole:2006p41}; and the substructure fraction (\fsub), which we take here to be the ratio of the particle count of all but the most massive of a FoF halo's substructures to its total particle count.  Each have simple physical interpretations as measures of dynamical state.  Elevated values of \fsub\ naturally arise during the earliest stages of a merger when a halo is naturally split between multiple similarly sized substructures.  Elevated values of \Vir\ are found prior to the dissipation of orbital energy following a merger.  Lastly, elevated values of \xoff\ are a natural result of the movement of a halo's dense core as it orbits the centre of mass of its system following even minor disturbances.  The standard values for relaxed systems which we adopt are those proposed originally by \citet{Neto:2007p2556} and recently confirmed to be successful in the study of halo density profiles \citep{Ludlow:2014p2558}; specifically, \xoff${<}0.07$, \Vir${<}1.35$ and \fsub${<}0.1$.  To date, a careful examination of how these metrics evolve following dynamical disturbances has not been performed however, leaving the physical nature of these cuts unclear.  Additionally, it is unclear how appropriate they are for high-redshift studies.

In what follows we shall study the evolution of these relaxation metrics following three sorts of mass accretion event capable of driving dynamical disturbances: halo formation (defined as the point at which a halo last reached 50\% of its present mass; a standard choice in the field, with other similar choices resulting in no qualitative change to our results), mergers between a primary halo and a secondary halo at least one third its mass (so-called `major', or 3:1 mergers) or mergers between a primary halo and a secondary halo at least one tenth its mass (so-called `minor', or 10:1 mergers).  Throughout our analysis we will measure intervals of time for a halo at redshift $z$ in units of its dynamical time, which we take to be 10\% of the Hubble time at that redshift.  Times in this dimensionless system of units will be denoted by $\tau$.  At all redshifts, the Hubble time is $\tau{=}10$ in this system.  The three times since a halo last experienced each of these events will be referred to as \emph{dynamical ages} and denoted by \tauform, \tauthreetoone\ and \tautentoone\ respectively.  The dynamical ages required for our relaxation criteria (\xoff, \Vir\ and \fsub) to return to and maintain standard values for relaxed halos following these events are used to motivate two \emph{recovery times}: a formation recovery time and a merger recovery time.  

\begin{figure*}
\begin{minipage}{170mm}
\begin{center}
\includegraphics[width=170mm]{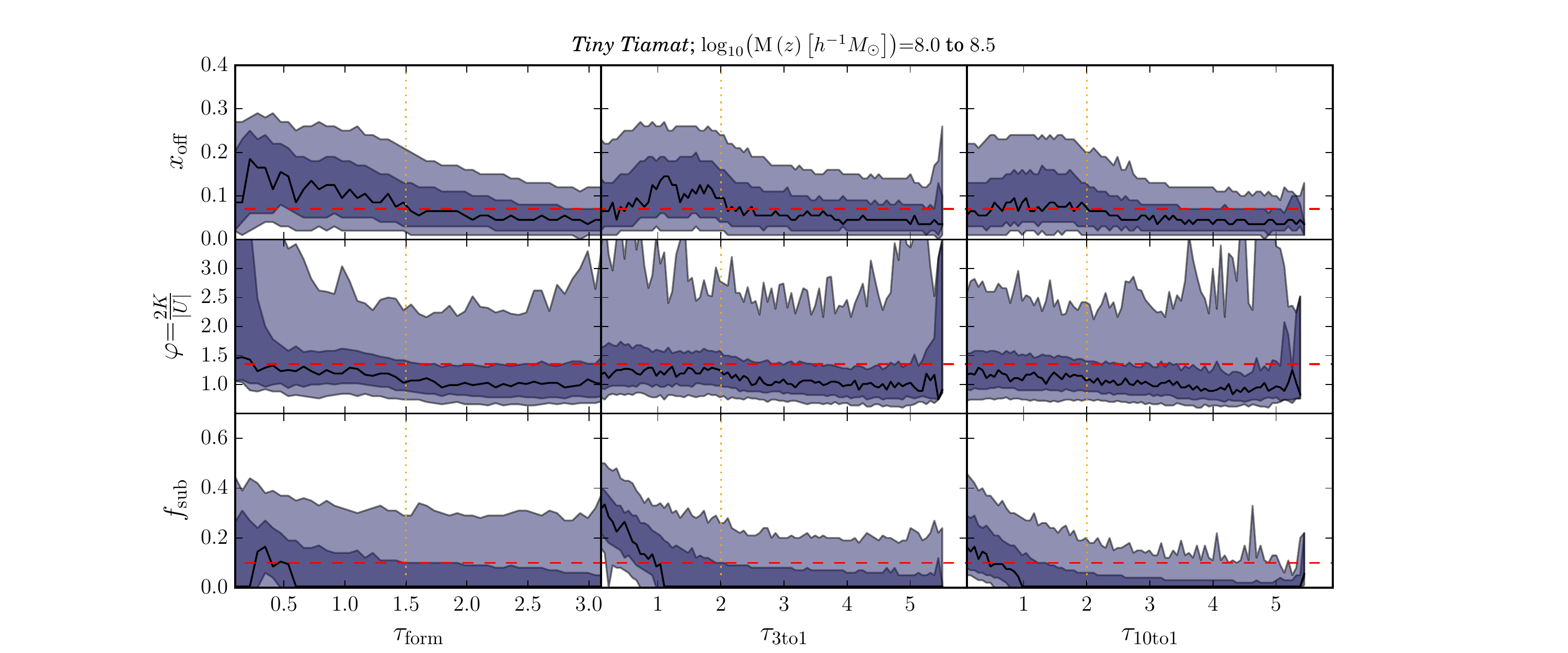}
\includegraphics[width=170mm]{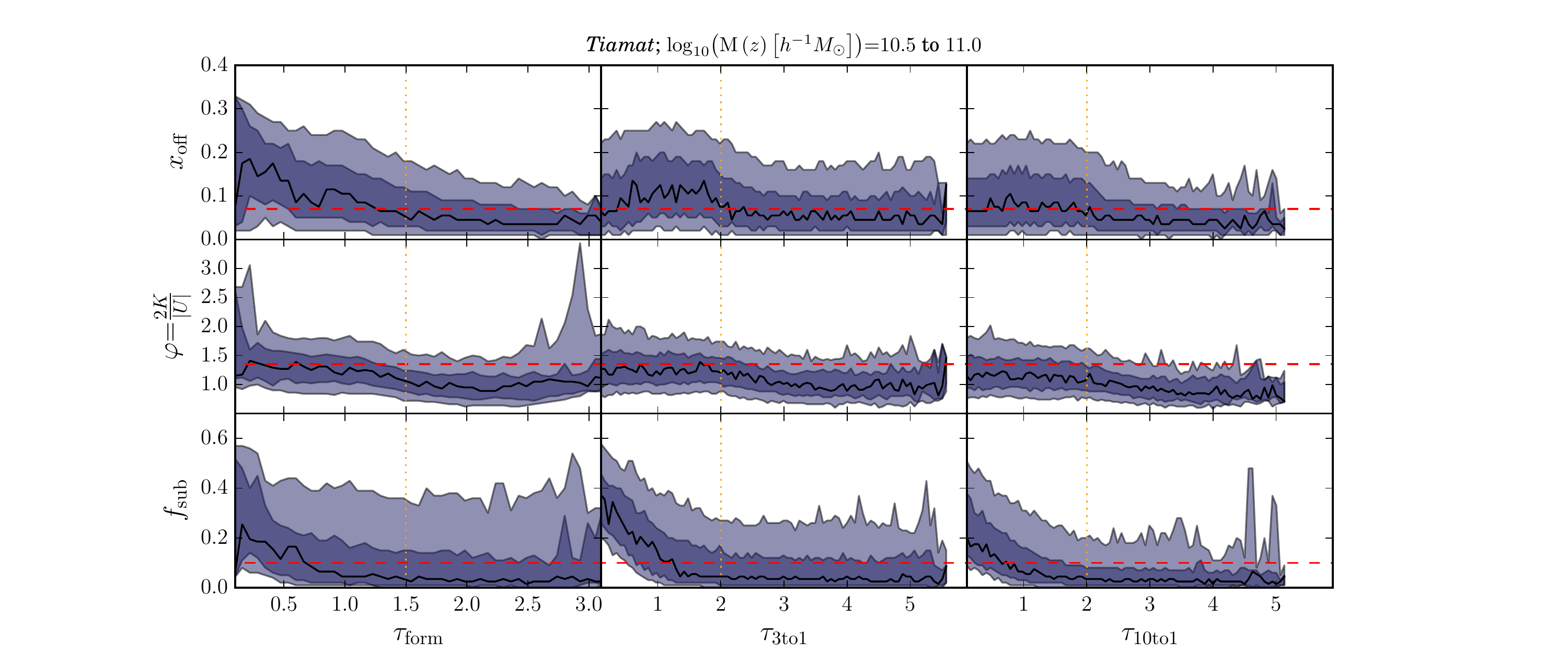}
\caption{Dependance of the offset parameter (\xoff), virial ratio (\Vir) and the substructure fraction (\fsub) with the time since a FoF halo's progenitor line: [left] achieved 50\% of its current mass (\tauform), [mid] last experienced a 3:1 (or larger) merger (\tauthreetoone) or [right] last experienced a 10:1 (or larger) merger (\tautentoone).  For a halo at redshift $z$, times are measured in units of its dynamical time, taken to be 10\% of the Hubble time at $z$ (in this system, the Hubble time is always $\tau{=}10$).  Black lines trace the peak of the distribution while dark and light shaded regions represent 68\% and 95\% confidence intervals about this peak respectively.  Standard relaxation criteria (\xoff${{=}0.07}$, \Vir${=}1.35$ and \fsub${{=}0.1}$) are labeled with dashed red lines and our fiducial recovery times (see text) are labeled with dotted orange lines.  All results are accumulated for halos over the redshift range $5{\le}z{\le}7.5$ for two masses (top, \Mvir${=}10^8{-}10^{8.5}$ \Msolhunit\ from \TinyTiamat\ and bottom, \Mvir${=}10^{10.5}{-}10^{11}$ \Msolhunit\ from \Tiamat) spanning a factor of 1000 in mass. \label{fig-recovery_Tiamat}}
\end{center}
\end{minipage}
\end{figure*}

In Figure \ref{fig-recovery_Tiamat} we show the evolution of the distribution of \xoff, \Vir\ and \fsub\ for two mass-selected halo populations\footnote{The mass ranges selected for Figure \ref{fig-recovery_Tiamat} were chosen because they yield sufficient statistics (575890 halos contribute to the \Tiamat\ panel) while consisting of halos large enough to resolve both the last 3:1 and 10:1 mergers in their progenitor lines.  A halo mass of $\log({\rm M}_{\rm vir}/{\rm M}_{\odot}){=}10.25$ consists of 6840 particles in \Tiamat.  The secondary halo of a 10:1 merger has only 684 particles in this case.  Given that these halos are doubling their mass every 1 to 2 dynamical times and we want to follow remnants for 3--4 dynamical times following merger events, these secondary halos could have been as small as 150 particles when they merged with their descendant.} (\Mvir${=}10^8{-}10^{8.5}$ \Msolhunit\ taken from \TinyTiamat\ and \Mvir${=}10^{10.5}{-}10^{11}$ \Msolhunit\ taken from \Tiamat) between $z{=}5$ and $z{=}7.5$ as a function of their three dynamical ages (we have looked at various redshift and mass ranges, finding no evidence of changes to any reported trends).  General trends are quickly apparent for each metric as a function of all three dynamical ages.  Since different environments, merger orbital properties and the oscillatory nature of \xoff\ and \Vir\ \citep[see][for an analysis]{Poole:2006p41} all lead to scatter in each metric as a function of $\tau$, we focus our discussion here on the path traced by the distribution peak of each metric, unless stated otherwise.  

We find that \xoff\ starts with high values of ${\sim}0.2$ at the time of halo formation, declining to our relaxed level of $0.07$ at \tauform${\sim}1.5$ and then to baseline levels of \xoff${\sim}0.04$ afterwards.  Following 3:1 and 10:1 mergers, peak levels occur roughly one dynamical time after a merger begins with relaxed levels obtained at \tauthreetoone${\sim}$\tautentoone${\sim}$2.  Peak values of $0.2$ and roughly $0.07$ are reached following 3:1 and 10:1 mergers, suggesting that mergers are progressively less likely to excite the system above our \xoff${\sim}0.07$ relaxation criterion as mass ratios drop below 10\%.

Interestingly, the virial ratio shows significantly less evolution following both formation and merger events.  In all cases, the distribution peak sits at levels similar to our \Vir${\sim}1.35$ relaxation criteria at times when \xoff\ lies above its relaxation criteria.  Once \xoff\ is found to drop below relaxation levels (or shortly before) \Vir\ can be seen to decline somewhat from values of ${\sim}$1.35 to ${\sim}$1.  Interestingly, the lower mass halos from the \TinyTiamat\ simulation behave similarly to the higher mass halos in \Tiamat, although with a significantly higher tail in the 95\% confidence interval.  Generally however, \Vir\ exhibits much less sensitivity to dynamical disturbances and relaxes to baseline levels quicker than \xoff, suggesting that it is a much less robust discriminator of dynamical state.

Lastly, \fsub\ shows a very simple and well defined behaviour following dynamical events.  At formation, a wide range of values are seen about a distribution peak of ${\sim}$0.2.  A slow decline to baseline values follows.  After merger events, \fsub\ increases by expected amounts: 30\% for 3:1 mergers and 10\% for 10:1 mergers.  The subsequent decline in \fsub\ is more rapid than what is seen following formation, with levels dropping at a rate of approximately 20\% per dynamical time.  We also find that substructure fractions return to standard relaxed values 30 to 50\% faster than core offsets following dynamical disturbances.  Despite this, because \fsub\ is most sensitive to dynamical disturbance in the earliest stages of mergers, it is an effective compliment to the \xoff\ statistic which exhibits a slight delay in reacting during merger events.

\begin{figure*}
\begin{minipage}{170mm}
\begin{center}
\includegraphics[width=170mm]{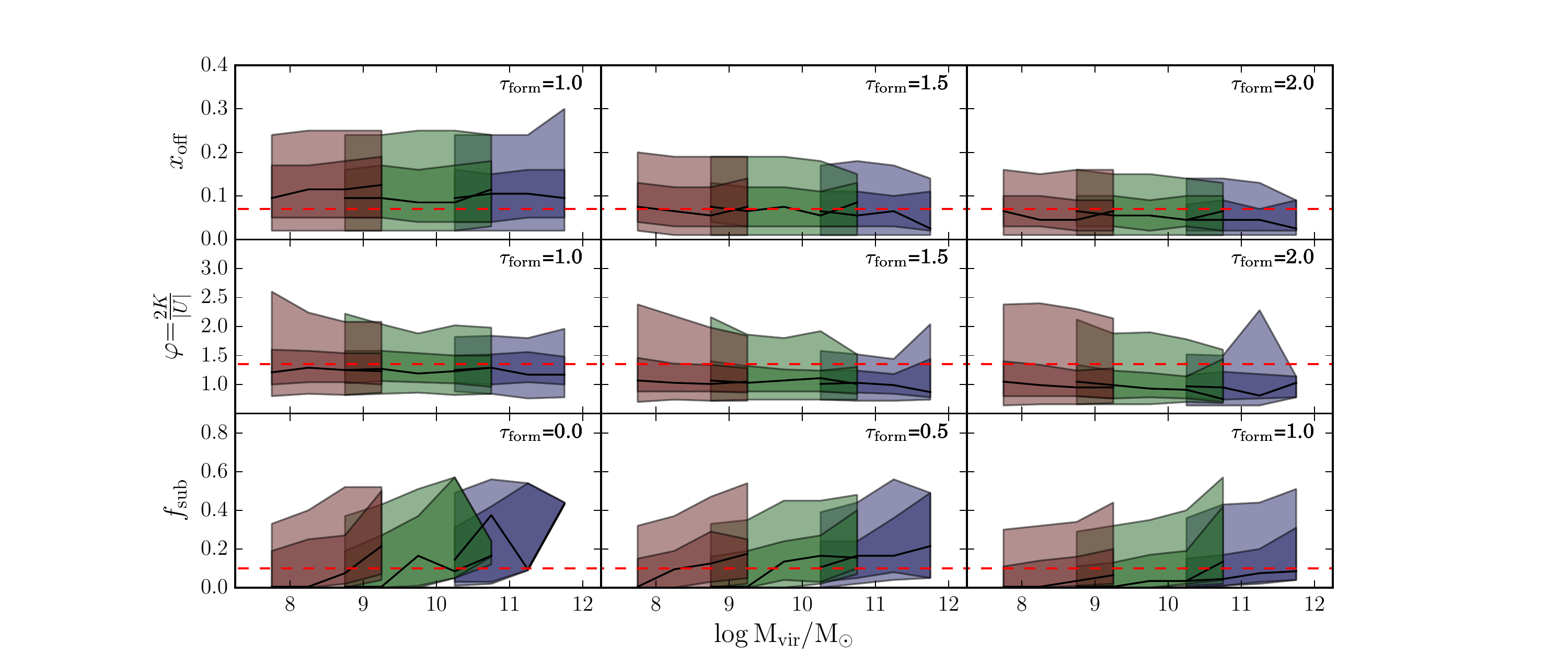}
\includegraphics[width=170mm]{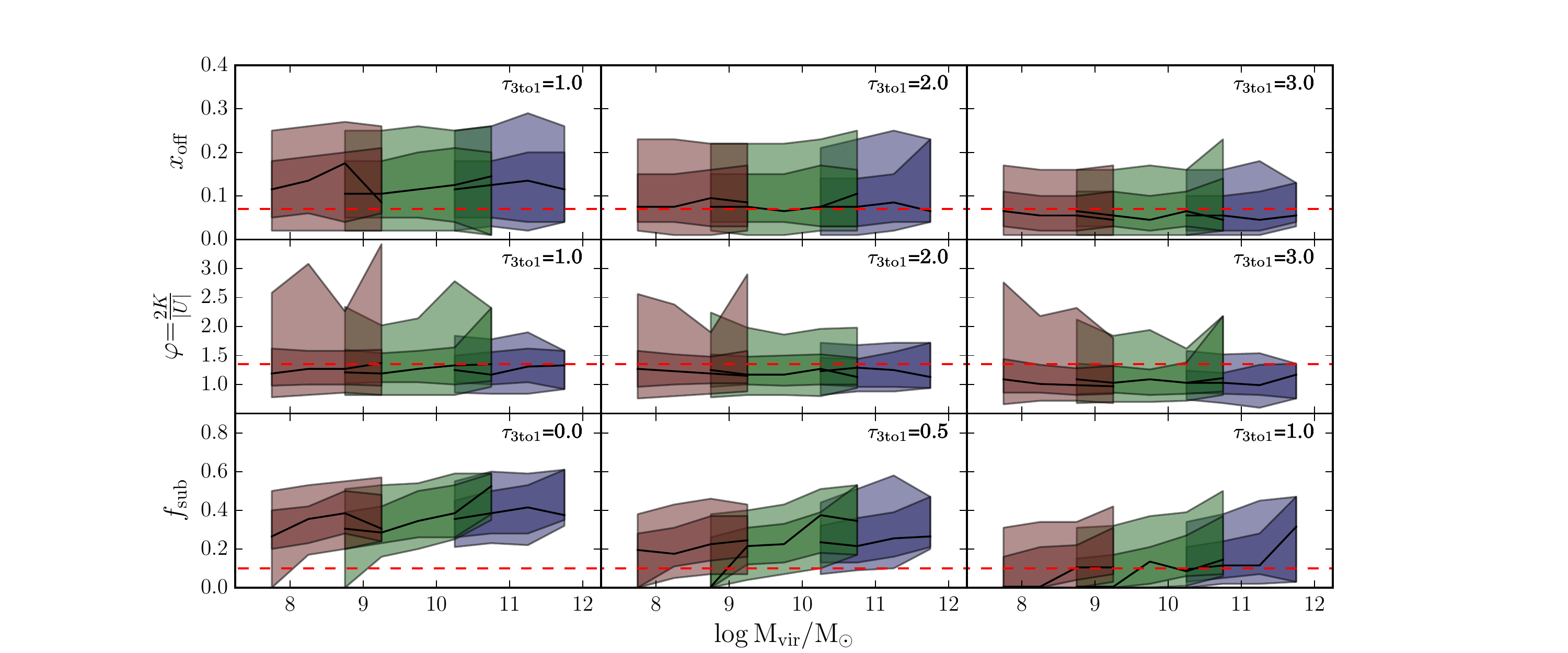}
\includegraphics[width=170mm]{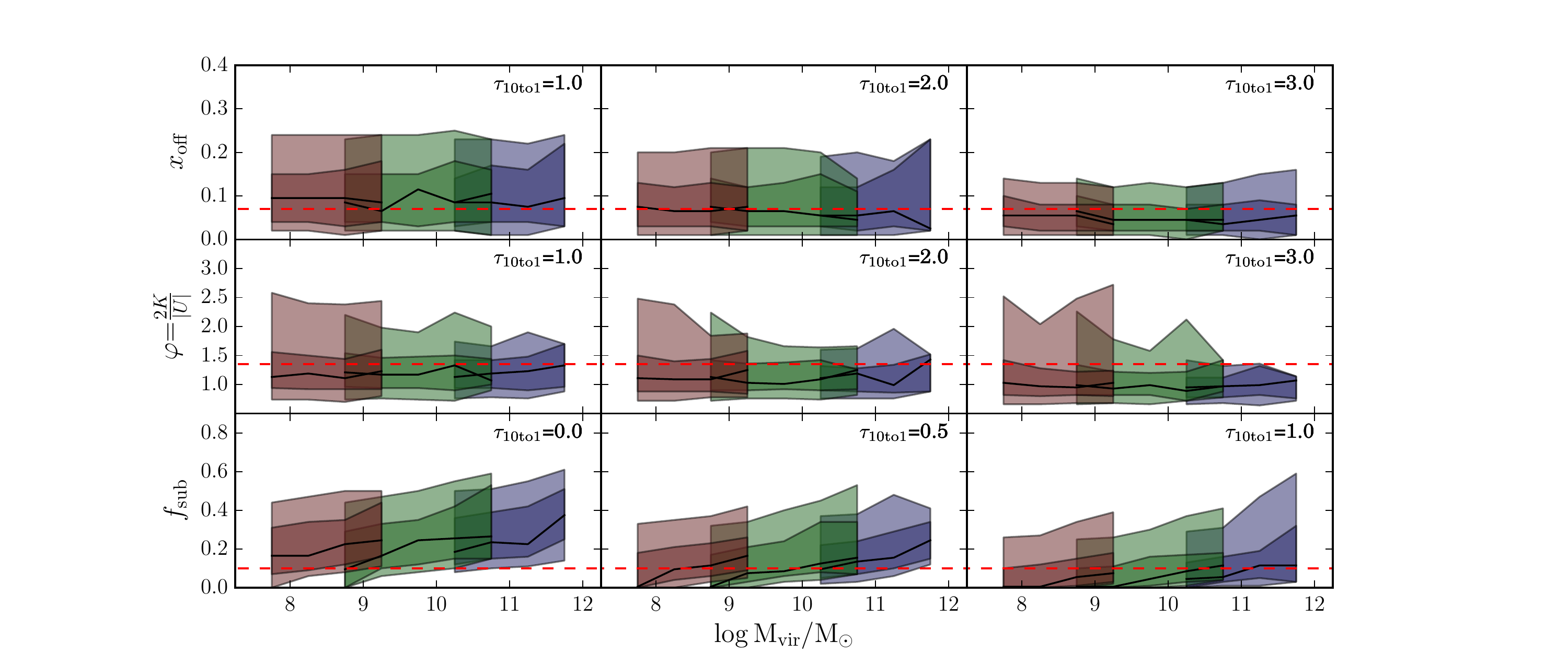}
\caption{Dependance of the offset parameter (\xoff), virial ratio (\Vir) and the substructure fraction (\fsub) on halo mass at a series of fiducial times measured since a FoF halo's progenitor line: [top] achieved 50\% of its current mass (\tauform), [mid] last experienced a 3:1 (or larger) merger (\tauthreetoone) or [bottom] last experienced a 10:1 (or larger) merger (\tautentoone).  For a halo at redshift $z$, times are measured in units of its dynamical time, taken to be 10\% of the Hubble time at $z$ (in this system, the Hubble time is always $\tau{=}10$).  Black lines trace the peak of the distribution while dark and light shaded regions represent 68\% and 95\% confidence intervals about this peak respectively.    Red, green and blue represents results taken from the \TinyTiamat, \MediTiamat, and \Tiamat\ simulations respectively.  Standard relaxation criteria (\xoff${{=}0.07}$, \Vir${=}1.35$ and \fsub${{=}0.1}$) are labeled with dashed red lines.  All results are accumulated for halos over the redshift range $5{\le}z{\le}7.5$.  The fact that each frame is essentially independent of mass (with the exception of \fsub, which is tilted across the range of each simulation, as expected, due to resolution effects) reflects the mass independence of how each metric relaxes following halo formation or mergers more significant than 10:1. \label{fig-recovery_Tiamat_M}}
\end{center}
\end{minipage}
\end{figure*}

We conclude then that, of the three metrics we study here, the \xoff\ statistic is the most effective single measure of dynamical state.  It is sensitive to disturbances from mergers greater than approximately 10:1 and retains this sensitivity for approximately 2 dynamical times afterwards.  Of course, this quantity is expected to oscillate following dynamical disturbances and shows a delay in responding to merger events, making the joint application of a complimentary and (ideally) independent metric necessary.  The \fsub\ metric is effective in this regard but looses sensitivity at late times when \xoff\ continues to maintain elevated levels.  The virial ratio is significantly less discriminating than these statistics but evolves in ways consistent with the relaxation of \xoff\ and \fsub.  Additional details (including specific numbers) regarding the relative influence of each statistic on setting relaxed halo population sizes in \Tiamat\ can be found in Angel et al. (2015; PAPER-II).

From Figure \ref{fig-recovery_Tiamat} we also note the remarkable similarity between the relaxation evolution of these two halo populations for all three metrics despite spanning a range of 1000 in mass.  In Figure \ref{fig-recovery_Tiamat_M} we present the distribution of all three metrics at three fixed times spanning the important range of their relaxation following each of the three dynamical events presented in Figure \ref{fig-recovery_Tiamat}.  With the exception of \fsub, all metrics are essentially independent of mass throughout the period of relaxation following halo formation or mergers larger than 10:1.  The differing trends of \fsub\ with mass for each simulation is a numerical effect arising from their differing resolutions as a function of mass and is an expected result.  Despite this one numerical effect, this figure clearly illustrates that high-redshift halos recover from formation and merger events within a time which is highly insensitive to their mass.

These results suggest that following formation or mergers greater than 10:1, a small and fixed number of pericentric passages of the material disturbed at large radius in the merger remnant are required for relaxation (many more passages may be involved at small radii where densities are higher and dynamical times shorter).  If this is the case, the mass independence of relaxation could be seen as a product of the fact that halo crossing times depend only on their mean density, which is defined in terms of a fixed overdensity, and independent of mass.  Secondary factors which could influence halo relaxation include halo concentrations, shapes and merger orbital properties.  While we have not yet been able to explore the mass dependance of halo shape and merger orbital properties and any possible influence they may exert on halo relaxation, we show in the next paper in this series (Angel et al. 2015; PAPER-II) that halo concentrations are nearly mass-invariant at $z{>}5$.  This may be partially responsible for the mass invariance of halo relaxation at high redshift in a way that may break down at low redshifts where halo concentrations do in fact have a significant mass dependance.

From these results, we define two mass-independent recovery times separating relaxed and unrelaxed systems at high redshift: \tauform${=}1.5$ and \taumerge${=}2$.  Our expectation is that high-redshift halos which have doubled their mass within one and a half dynamical times (\tauform${<}1.5$) or which have experienced mergers larger than 10:1 within two dynamical times (\taumerge${<}2$) are likely to be disturbed.

\subsection{Relaxed fraction evolution}\label{sec-relaxation_evolution}

\begin{figure*}
\begin{minipage}{170mm}
\begin{center}
\includegraphics[width=170mm]{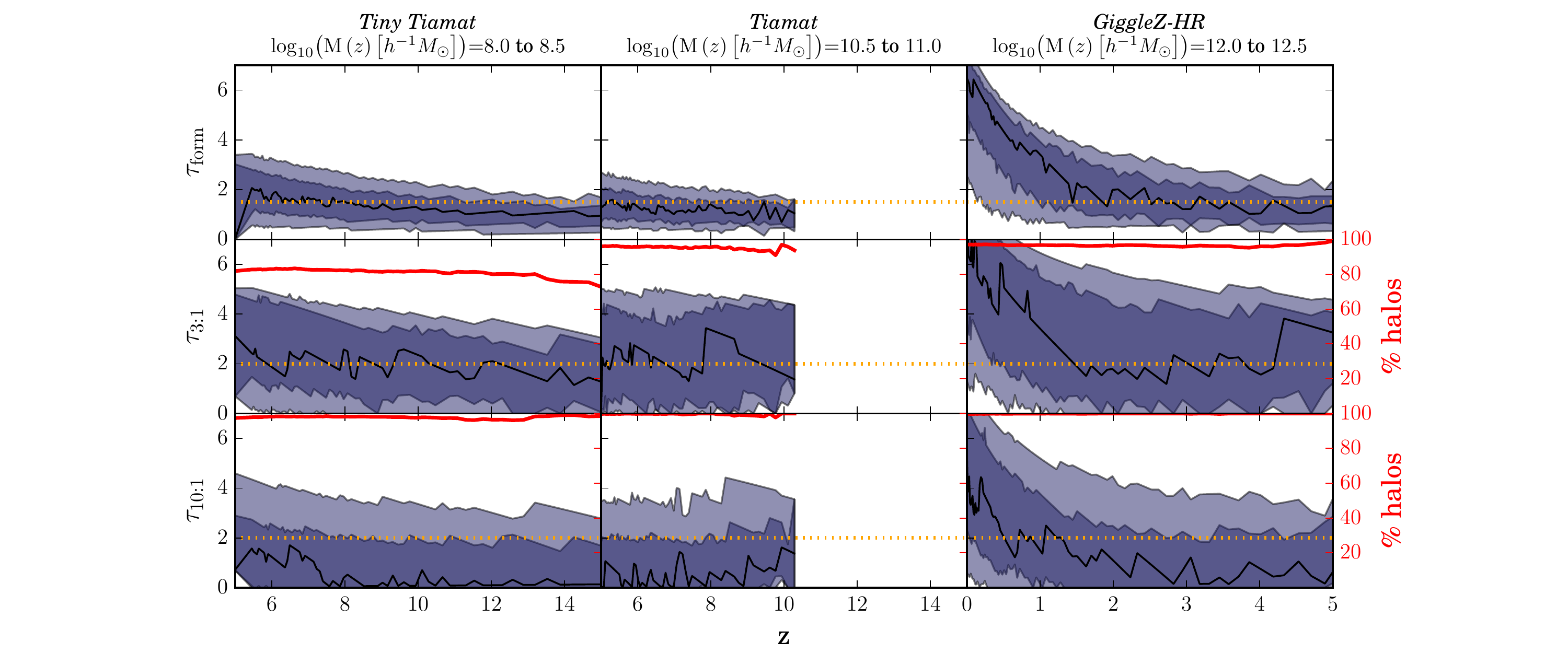}
\caption{Redshift dependance of the dynamical age (black lines show the distribution peak while dark and light shaded regions represent 68\% and 95\% confidence intervals about this peak respectively) of FoF halos for two cases at $z{>}5$ spanning a range of 1000 in mass ([left] \Mvir${=}10^8{-}10^{8.5}$ \Msolhunit\ from \TinyTiamat\ and [middle] \Mvir${=}10^{10.5}{-}10^{11}$ \Msolhunit\ from \Tiamat) contrasted with [right] the $z{<}5$ evolution of \Mvir${=}10^{12}{-}10^{12.5}$ FoF halos from the \GiggleZHR\ simulation.  Our fiducial recovery times (\tauform${=}1.5$ and \taumerge${=}2$; halos above these lines are capable of being relaxed) are labeled with dotted orange lines and the fraction of halos that have experienced 3:1 or 10:1 mergers are shown in red.
\label{fig-tau_z_plots}}
\end{center}
\end{minipage}
\end{figure*}

How then do the fractions of halos meeting these recovery criteria evolve with redshift?  In Figure \ref{fig-tau_z_plots} we plot the evolution of the distributions of \tauform,\tauthreetoone, and \tautentoone\ as a function of redshift for the two mass-selected populations presented in Figure \ref{fig-recovery_Tiamat}.  The distributions of dynamical ages are shown in blue (black lines show the distribution peak while dark and light shaded regions represent 68\% and 95\% confidence intervals about this peak respectively) and the fraction of the population which have had 3:1 or 10:1 mergers are plotted in red (nearly 100\% in all cases).  The \tauform\ and \taumerge\ recovery times obtained in Section \ref{sec-recovery_timescales} are indicated with dotted orange lines.  Our expectation is that halos existing above the indicated recovery times for all three cases should be relaxed.

\begin{figure*}
\begin{minipage}{170mm}
\begin{center}
\includegraphics[width=170mm]{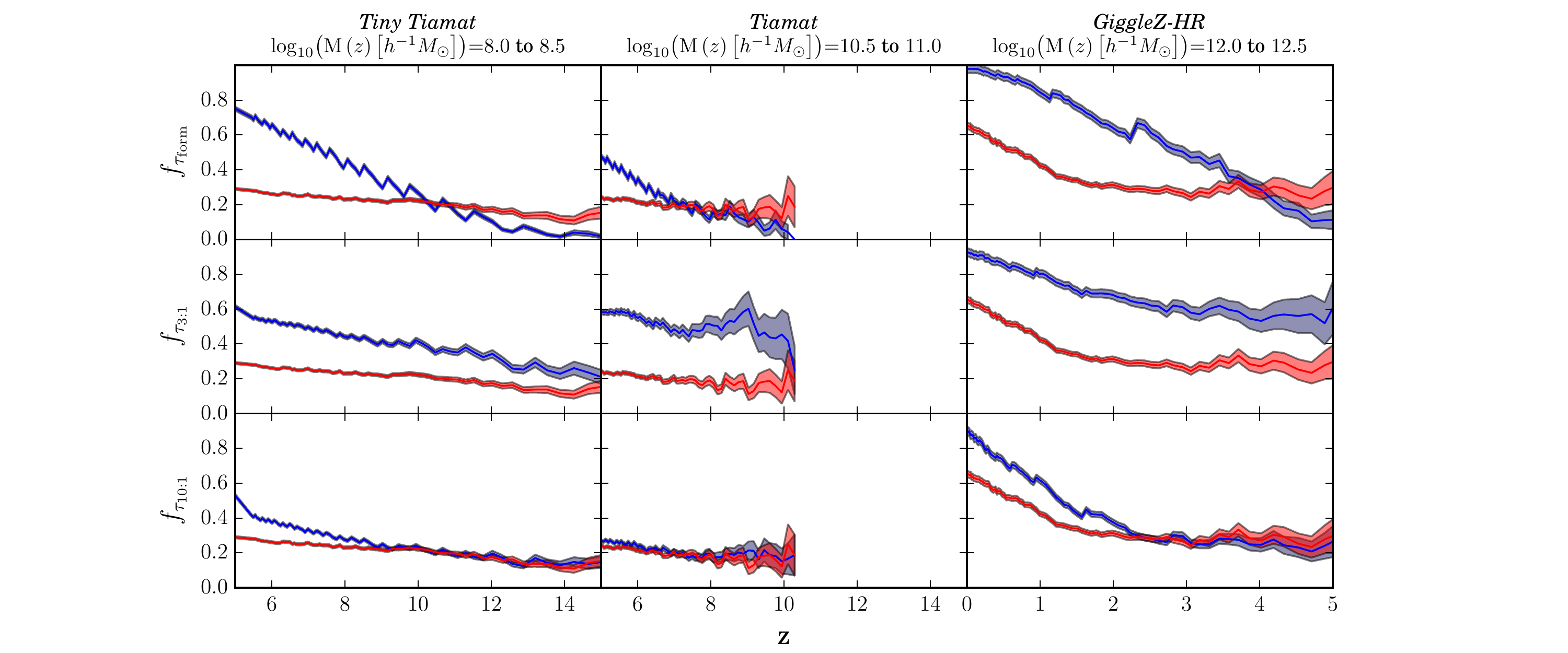}
\caption{Redshift dependance of the fraction of FoF halos meeting our recovery (in blue) and relaxation (in red) criteria (black lines show the fraction of the population while shaded regions depict poisson uncertainties) for two cases at $z{>}5$ spanning a range of 1000 in mass ([left] \Mvir${=}10^8{-}10^{8.5}$ \Msolhunit\ from \TinyTiamat\ and [middle] \Mvir${=}10^{10.5}{-}10^{11}$ \Msolhunit\ from \Tiamat) contrasted with [right] the $z{<}5$ evolution of \Mvir${=}10^{12}{-}10^{12.5}$ FoF halos from the \GiggleZHR\ simulation.  In all cases, the recovered fraction roughly follows the fraction of halos that have been able to recover from their last 10:1 (or larger) merger, suggesting that minor mergers are principally responsible for regulating the relaxed fraction of a halo population.  Deviations between the two at $z{<}2$, when growth stops being exponential (indicated by \tauform${\sim}$constant), may reflect important changes at low redshift to the processes of dynamical recovery.\label{fig-recovered_fractions}}
\end{center}
\end{minipage}
\end{figure*}

We can see from this figure that the distribution of all three dynamical ages evolves very little from $z{=}15$ to $z{=}5$.  Over this redshift range, the distribution of formation ages is very narrow and peaked very close to our formation recovery timescale of \tauform${=}1.5$.  The near constant value of \tauform\ during this epoch is consistent with early mass accretion histories which are exponential, as found previously by several other authors \citep[\eg][]{Wechsler:2002p1783,McBride:2009p1230,Correa:2015p2654}.  The distribution of times since 3:1 mergers is much broader and is also peaked near our merger recovery timescale \taumerge${=}2$.  This tells us that typical halos at high redshifts across all galactic masses are doubling their mass on timeframes that only barely permit relaxation while simultaneously, major mergers are occurring at rates which only barely permit recovery between events.  The situation is importantly different for minor mergers.  In this case we find that halos experience minor mergers at rates which are much too rapid (on average) to permit dynamical relaxation between events. 

In Figure \ref{fig-tau_z_plots} we also plot the low-redshift evolution of our three dynamical ages as found in the \GiggleZHR\ simulation. We note that the distributions compare well between \Tiamat\ and \GiggleZHR\ at $z{=}5$ despite slightly different cosmologies and the somewhat higher masses depicted by \GiggleZHR, validating its use for qualitative comparisons.  Here we find that the narrow distribution of formation ages, broad distribution of merger ages and short times between 10:1 mergers persists almost unchanged until approximately $z{\sim}2$.  At this time, we find that \Mvir${=}10^{12}{-}10^{12.5}$ \Msolhunit\ halos begin to become progressively older, as typical formation ages and times since major mergers increase and times since minor mergers creep above merger recovery times of \taumerge${=}2$ by $z{=}0$.  The increase of \tauform\ at lower redshifts corresponds to a transition in these halos' mass accretion histories from an exponential form to a linear form, as discussed already in the literature \citep[\eg][]{McBride:2009p1230,Correa:2015p2654}.

The fractions of halos which meet these recovery criteria as a function of redshift is presented explicitly in Figure \ref{fig-recovered_fractions}.  Here we see more clearly (in blue) the trends we identified from Figure \ref{fig-tau_z_plots}: the disappearance of halos with formation times less than \tauform${=}1.5$ and the sustained low levels of halos having had sufficient time to recover from their most recent mergers.  We have added to these plots (in red) the fraction of halos that simultaneously satisfy our standard \xoff, \Vir,and \fsub\ relaxation criteria.  Remarkably, the fraction of relaxed halos and the fraction having had sufficient time to recover from their last 10:1 (or larger) merger are very similar across a wide range of masses and redshifts.  We conclude from this that the standardised relaxation criteria of \citet{Neto:2007p2556} are effectively identifying systems that have been disturbed by 10:1 (or larger) mergers.  It should be noted however that our recovery criteria of \tauform${=}1.5$ and \taumerge${=}2$ have been calibrated at high redshift and may need adjustment at low redshift, where halos are substantially more concentrated and the orbital properties of merging systems are significantly different, with more circular orbits requiring longer to relax.  This is likely the reason why our estimates of the recovered fraction (measured using criteria calibrated at $z{>}5$) exceeds the relaxed population (as measured directly from \xoff,\Vir\ and \fsub) at $z{<}2$.

Taken together, we see that at high redshifts ($z{>}5$), the fraction of relaxed halos drops to levels of ${\sim}20$\% at all galactic masses.  Combined with the rapid decline in the number density of halos with redshift at this time, we conclude that the abundance of relaxed galactic halos prior to the epoch of reionization drops to very low levels.  This should make it very challenging to assemble large populations of relaxed halos at $z{>}10$, which is of particular concern for studies seeking to understand the processes acting to establish universal density profiles for collisionless systems at high redshift.
 
\subsection{Large and long-lived phase-space structures}

\begin{figure*}
\begin{minipage}{170mm}
\begin{center}
\includegraphics[width=80mm]{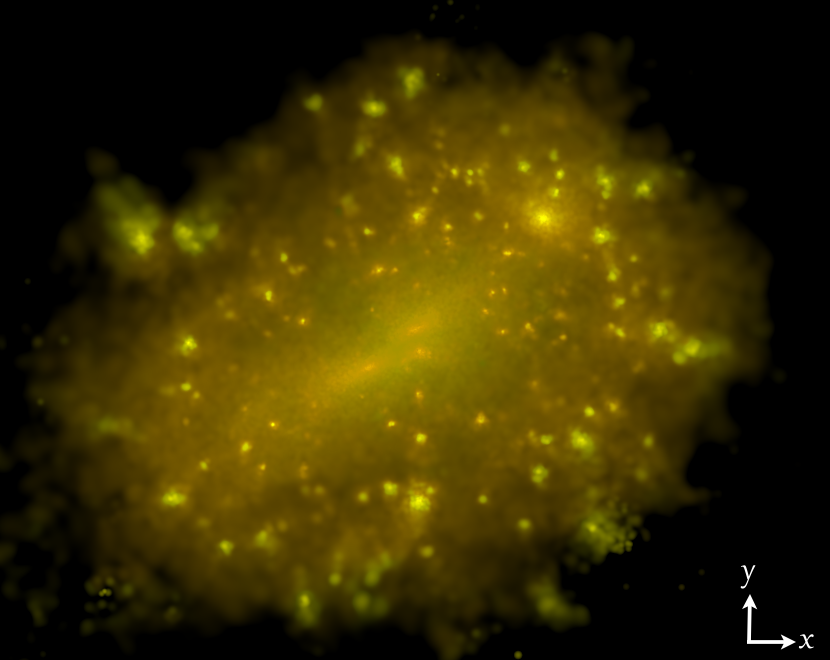}
\includegraphics[width=80mm]{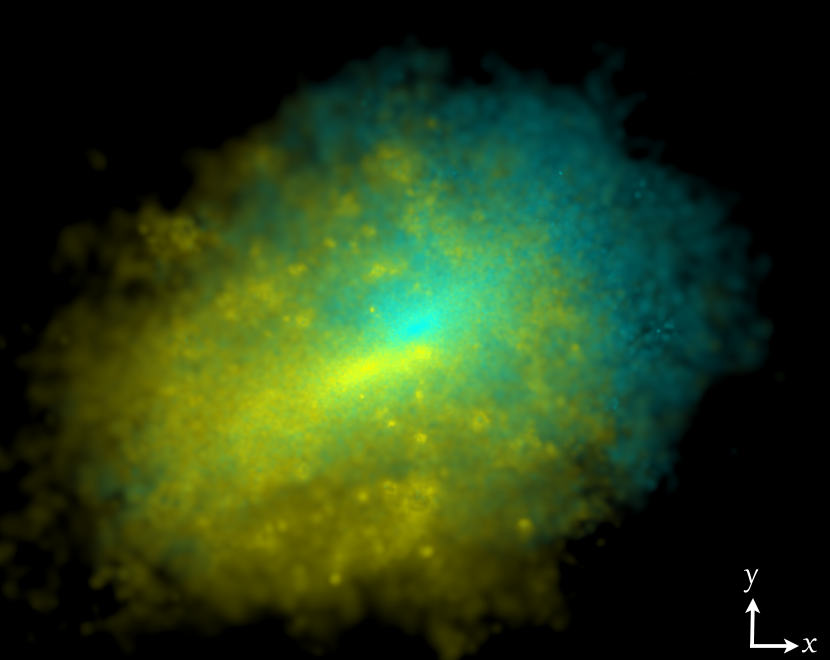}~\includegraphics[width=80mm]{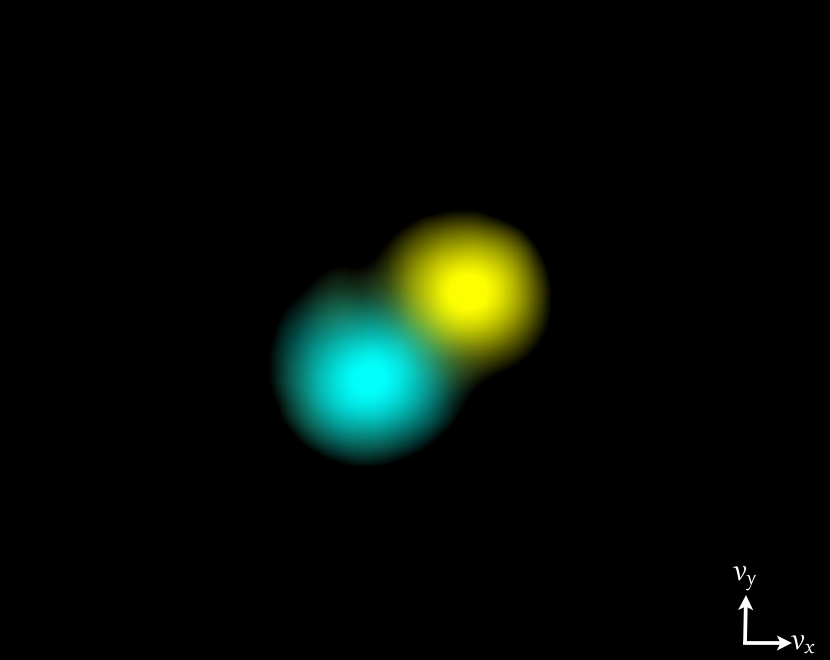}
\includegraphics[width=80mm]{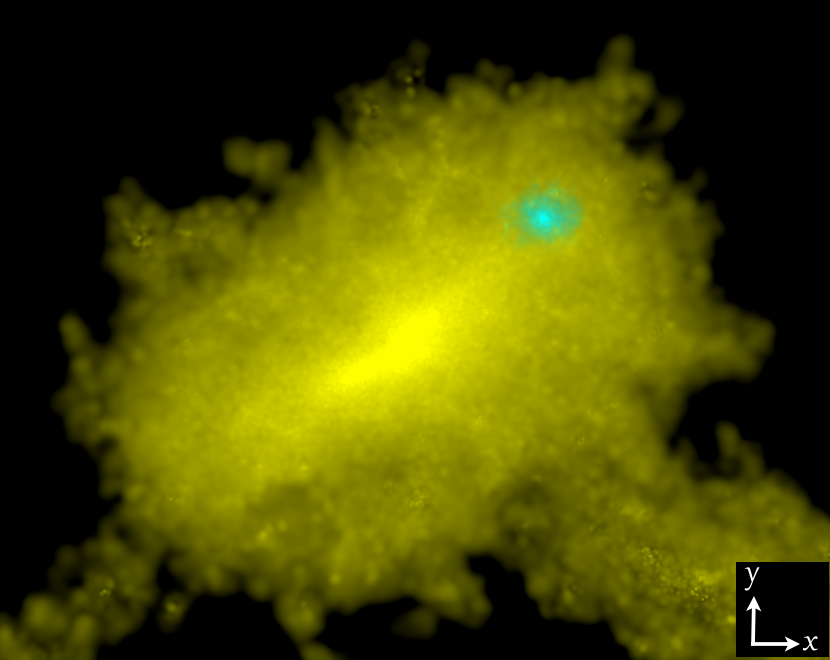}~\includegraphics[width=80mm]{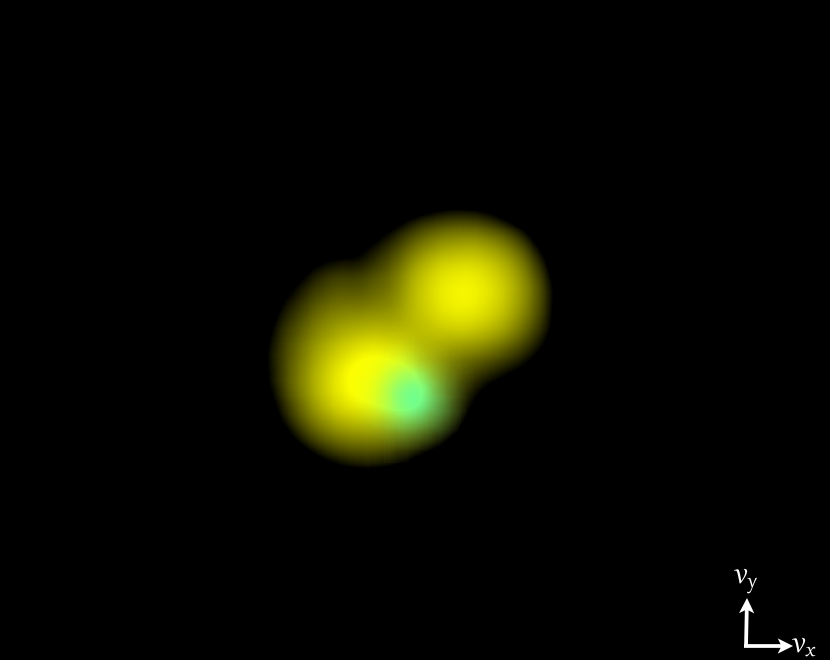}
\caption[Example of a massive Tiny Tiamat halo]{A \Mvir${=}4.4{\times}10^{10}$\Msolhunit\ halo at $z{=}5$ in the \TinyTiamatW\ simulation.  The top panel is a configuration-space rendering of the FoF halo as identified by \ROCKSTAR\ with colours set by the column weighted particle velocity dispersion.  The bottom four panels depict the [left] real-space and [right] velocity-space structure of the two most massive substructures (yellow for the most massive, cyan for the second most massive) identified in this FoF halo by [middle] \ROCKSTAR\ and [bottom] \Subfind.  Luminance is set by the logarithm of the integrated column through the system in all cases.  In the case of the cyan structure in the \Subfind\ case, luminance has been arbitrarily increased by a factor of 10 to increase its contrast.  Fields of view for all configuration and velocity space images are 200\kpchunit\ (comoving) and 1000\kms\ respectively.  Large distinct substructures such as the ones identified here by \ROCKSTAR\ are common and long lived at high redshifts.\label{fig-halo_example}} 
\end{center}
\end{minipage}
\end{figure*}

\begin{figure*}
\begin{minipage}{170mm}
\begin{center}
\includegraphics[width=170mm]{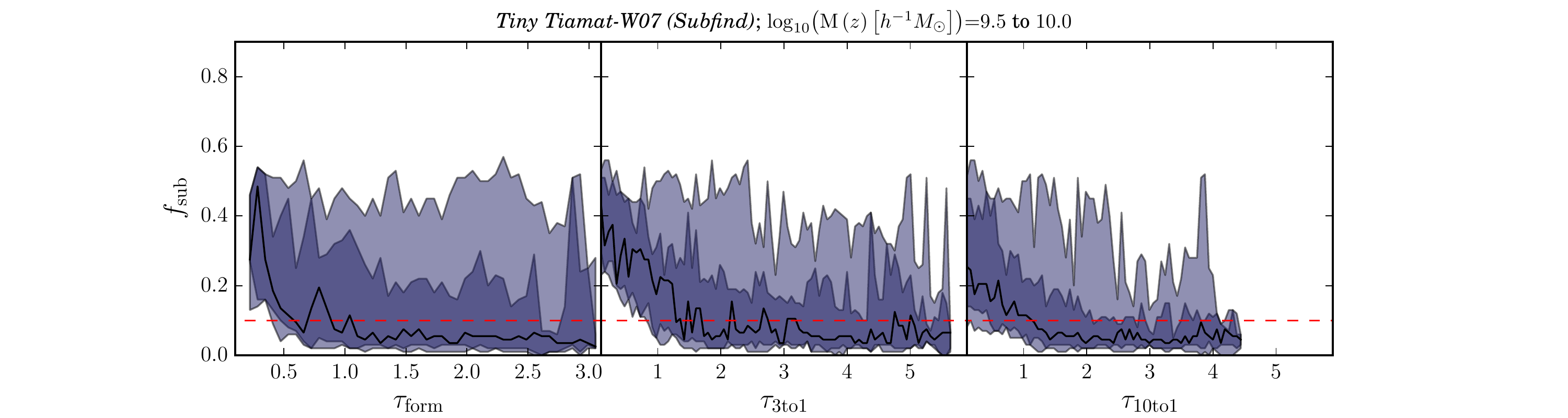}
\includegraphics[width=170mm]{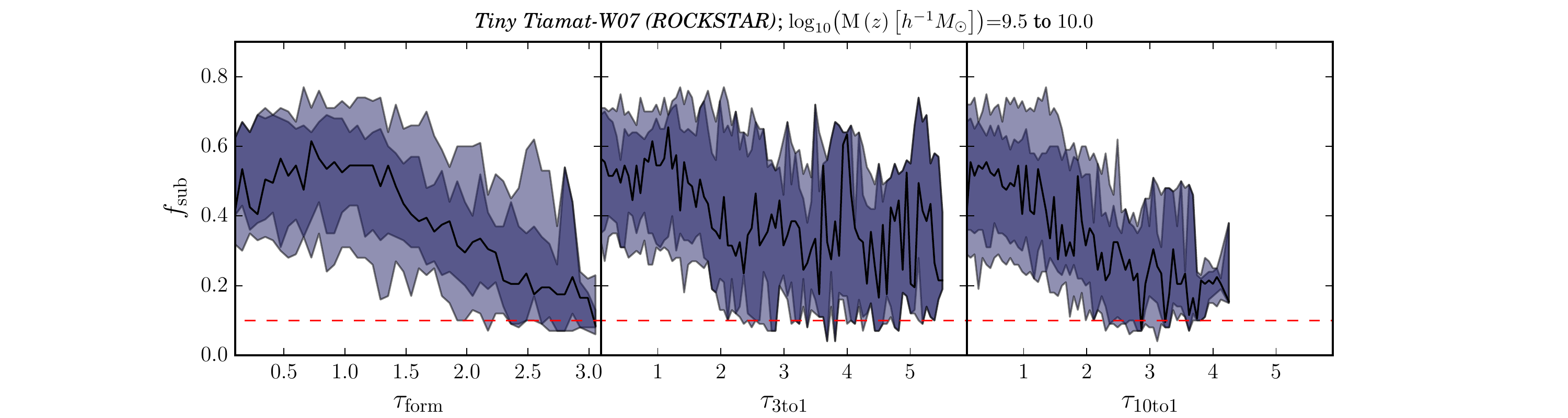}
\caption{Comparison of the evolution of substructure fractions (\fsub) derived from [top] \Subfind\ and [bottom] \ROCKSTAR, in a well resolved and populated mass range in the \TinyTiamatW\ simulation as a function of time since a FoF halo's progenitor line: [left] achieved 50\% of its current mass (\tauform), [mid] last experienced a 3:1 (or larger) merger (\tauthreetoone) or [right] last experienced a 10:1 (or larger) merger (\tautentoone).  Black lines show the distribution peak while dark and light shaded regions represent 68\% and 95\% confidence intervals about this peak respectively.  For a halo at redshift $z$, times are measured in units of its dynamical time, taken to be 10\% of the Hubble time at $z$ (in this system, the Hubble time is always $\tau{=}10$).  \ROCKSTAR\ halos have much higher substructure fractions due to the longer recovery times required for major mergers to loose their identity in phase-space as opposed to the shorter times required to loose their identity in configuration space. \label{fig-recovery_subfind_vs_rockstar}}
\end{center}
\end{minipage}
\end{figure*}

As an exercise during the development phase of the \Tiamat\ simulations, we analysed one of our simulations (\TinyTiamatW) with the \ROCKSTAR\ halo finder, allowing us to study the effect of halo finding on our semi-analytic modelling campaign.  Doing so has yielded an interesting new insight into the dynamical lives of high-redshift galactic halos.

In Figure \ref{fig-halo_example} we depict a relatively massive but otherwise typical halo at $z{=}5$ extracted from the \TinyTiamatW\ simulation.  In this figure, the whole FoF system (as reported by \ROCKSTAR) is depicted in the top panel while subsequent panels depict the configuration-space (left) and velocity-space structure (right) of the most massive substructure (in yellow) and the second most massive substructure (in cyan) as determined by \Subfind\ (middle) and \ROCKSTAR\ (bottom).  There is a stark difference between the results from these two halo finders.  We can see clearly -- in a manner we find to be shared by all massive halos at high redshift -- that while the halo appears relatively undisturbed with unremarkable substructure,  it in fact consists primarily of two very massive subhalos which are distinct in phase space.

Phase-space halo finders such as \ROCKSTAR\ are of course designed to separate halo substructures in this way, but it is not entirely clear that this is a desired result for applications in galaxy formation modelling.  While approaches differ in detail, the central premise of all semi-analytic galaxy formation models is that the total matter assembly provided by their merger tree inputs can be reliably mapped to a faithful description of the baryonic assembly of galactic halos.  Problems may arise if the collisional fluids (particularly the hot halos) associated with multiple collisionless systems oscillating through each other for $>$3 dynamical times can not follow the collisionless material of their initial hosts.  Substantial amounts of this gas will be stripped or rapidly coalesce into one hot halo, loosing its association with its original collisionless component while that material continues to orbit.  This is the case with the Bullet Cluster for instance, albeit at a different mass scale and redshift.  It is also the situation modelled by \citet{McCarthy:2008p2661} who find that the stripping of a galaxy's hot halo (due to tides, ram pressure stripping, and hydrodynamic instabilities) is extremely efficient up to and during its first pericentric passage (\ie\ $\sim1$ dynamical time following accretion).  The amount of material removed varies with halo mass, concentration and orbit, but is substantial and typically in the range of 60 to 80\% for the broad range of cases they examine.

If such structures were short lived, the impact on our galaxy formation model would likely be insignificant.  However, they are in fact long lived in dynamical terms.  Following the format of Figure \ref{fig-recovery_Tiamat}, Figure \ref{fig-recovery_subfind_vs_rockstar} presents a comparison of the evolving substructure fractions of FoF halos extracted from \TinyTiamatW\ using \Subfind\ to those obtained from \ROCKSTAR\ as functions of the dynamical ages \tauform,\tauthreetoone, and \tautentoone.  While we see the familiar decline of \fsub\ following formation and mergers presented in Figure \ref{fig-recovery_Tiamat} in the \Subfind\ trees, the \ROCKSTAR\ trees exhibit a much slower decline, reaching constant levels only after $\tau{\gtrsim}5$, sustaining levels well above our standard relaxation criteria even after that.

\begin{figure}
\includegraphics[width=80mm]{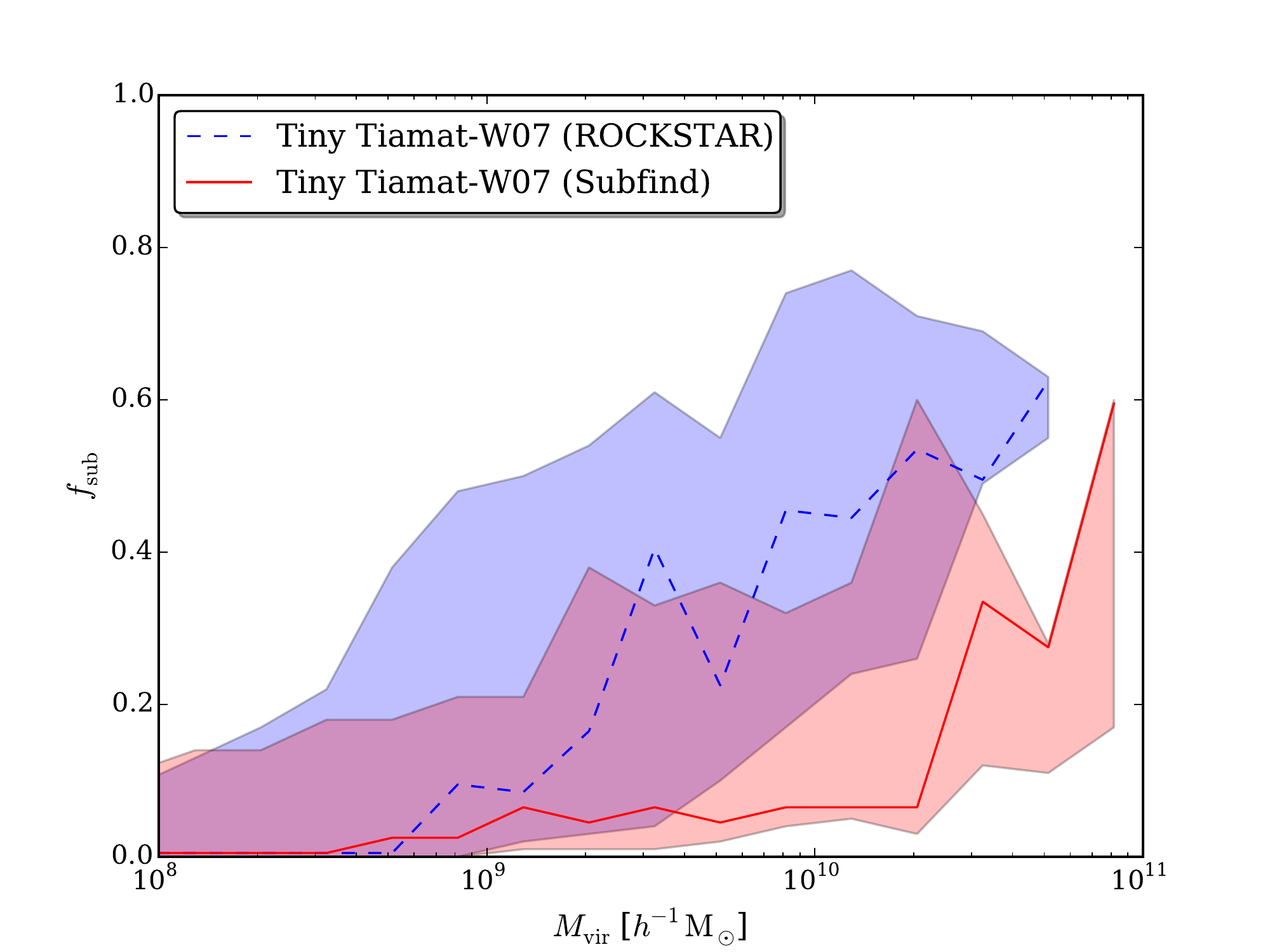}
\caption[Substructure functions]{Substructure fractions in the \TinyTiamatW\ simulation as measured by \ROCKSTAR\ (in blue) and \Subfind\ (in red) at $z{=}5$.  Lines (solid for \Subfind, dashed for \ROCKSTAR) depict the peak of the distribution while shaded regions depict the 68\% confidence interval about this peak.  The higher substructure fractions in \ROCKSTAR\ are a product of long lived phase-space structures resulting from major mergers which dissolve quickly (in a couple dynamical times) in configuration space but retain their identity in velocity-space.\label{fig-substructure_functions}} 
\end{figure}

On the other hand, if these substructures were rare, their impact on galaxy formation modelling would again be minimal.  They are in fact very common.  To illustrate their prevalence and to quantify the magnitude of this effect, we present the substructure fractions of our \Subfind\ and \ROCKSTAR\ $z{=}5$ \TinyTiamatW\ halo catalogs in Figure \ref{fig-substructure_functions} .  While the FoF halo mass functions for the two catalogs are virtually identical (except at the highest masses where the larger linking length used by \ROCKSTAR\ unsurprisingly yields more systems, presumably due to overlinking), the substructure fractions at the highest (and most resolved) masses of the two catalogs are very different.  Substructure fractions are 50 to 60\% at the highest masses in \ROCKSTAR\ indicating that only around half of the mass in these systems is assigned to the most massive component of the system.  This is a consequence of a very different splitting of the top level of the FoF group's substructure hierarchy.

Suggestions of this effect can be seen in the recent work of \citet{Behroozi:2015p2660}.  While these authors find that substructure properties like position and velocity generally agree between configuration and phase-space halo finders, they find that substantial differences in masses can occur.  They also find strong disagreements in the frequency and duration of major mergers, particularly at redshifts $z{>}1$.  These differences are likely related to the situation presented in Figure \ref{fig-halo_example}.

We emphasise that we make no attempt here to advocate for one halo finding approach over another.  Rather, we seek to make the point that care should be taken -- particularly at high redshift where major mergers are frequent and the sorts of large, diffuse phase-space structures we illustrate in Figure \ref{fig-halo_example} are likely most prevalent -- to ensure that each semi-analytic model is matched, in a physically meaningful way, to the nature of the substructure hierarchy supplied by the halo finder contributing to its input.  Such differences may lead (once tuneable parameters are adjusted to yield accurate fits to observations) to significant systematics with mass in the evolution of merger trees which could masquerade as physical processes as diverse as mass dependancies in dust properties, photon escape fractions, feedback and cooling.  A detailed account of how the cooling and feedback modelling of DRAGONS (using \Meraxes\ with trees derived from \Subfind\ halo finding) compares to the \Smaug\ hydrodynamic simulations of \citet{Duffy:2014p2561} will be presented in Qin \etal\ (2015, PAPER-VII), where a direct halo-by-halo comparison of the two methodologies will be presented.

We take this opportunity to point out one other possible important astrophysical consequence of large bulk phase-space structures such as this.  Recent studies have begun to investigate the possibility that heating from dark matter annihilation may be observable in the redshifted 21-cm background from $z{>}30$ \citep[\eg][]{Furlanetto:2006p2566,Evoli:2014p2655,Mack:2014p2565,Schon:2014p2564}.  If phase-space structures such as these prove to be common at this epoch, important changes to inferred annihilation cross sections may result.

%% file: summary.tex
We have introduced the Dark-ages Reionization and Galaxy-formation Observables from Numerical Simulations (DRAGONS) program and presented the \Tiamat\ collisionless N-body simulation suite upon which it is constructed.  The abundance of friends-of-friends (FoF) structures populating \Tiamat\ is a good match to the \quotes{universal} model proposed by \citet{Watson:2013p2555} at high masses, but we find a supression of low-mass systems, possibly due to differences in our halo finding procedure or perhaps indicating a deviation from \quotes{universal} behaviour, at least at large redshifts.

Using \Tiamat\ we have also illustrated the dynamically violent conditions experienced by galactic halos at large redshift.  We find that across a wide range of galactic mass (\Mvir${=}10^8$ to $10^{11}$ \Msolhunit) above $z{=}5$, halos relax from their formation and from mergers in essentially the same way and in the same amount of time: within one and a half dynamical times in the case of their formation (\ie\ \tauform${=}1.5$) and within two dynamical times following mergers involving a primary and a secondary larger than 10\% of its mass (\ie\ \taumerge${=}2$).

The distribution of formation times and times since major mergers maintain approximately these time-scales across all redshifts above $z{=}5$ while the time between minor mergers is typically significantly less.  Relaxed fractions maintain levels of less than 20\% at $z{>}5$ as a result.  Using the \GiggleZHR\ simulation (which extends to $z{=}0$, albeit with poorer resolution) we find that this remains true for \Mvir${=}10^{12}$ \Msolhunit\ halos until $z{\sim}2$.  It appears that the rate of minor mergers principally regulate a halo population's relaxed fraction, as measured by standard metrics.  Combined with the rapid decline of the halo mass function at redshifts $z{>}10$, the abundance of relaxed halos prior to the epoch of reionization must be extremely low.

Using the phase-space halo finder ROCKSTAR, we also demonstrate that high-redshift halos host large and long-lived substructures that go undetected to halo finders such as \Subfind\ which utilise configuration-space information only.  This results in substructure fractions that are much higher for \ROCKSTAR\ than for \Subfind, with probable implications for semi-analytic models of galaxy formation at high redshift.

Taken together, these results illustrate the dynamically violent circumstances under which galaxy formation proceeds in the early Universe.  The consequences are many and significant, including implications for photon escape fractions, efficiencies of feedback from winds (both stellar and AGN) and the efficiency of spheroid assembly.  These in turn can have important consequences for the reionization history of the Universe during the EoR and observed galaxy sizes.

%% file: thanks.tex
This research was supported by the Victorian Life Sciences Computation Initiative (VLSCI), grant ref. UOM0005, on its Peak Computing Facility hosted at the University of Melbourne, an initiative of the Victorian Government, Australia. Part of this work was performed on the gSTAR national facility at Swinburne University of Technology.  gSTAR is funded by Swinburne and the Australian Governments Education Investment Fund. This research program is funded by the Australian Research Council through the ARC Laureate Fellowship FL110100072 awarded to JSBW.  CP acknowledges support of ARC DP130100117 and DP140100198, and ARC Future Fellowship FT130100041.  AM acknowledges support from the European Research Council (ERC) under the European UnionÕs Horizon 2020 research and innovation program (grant agreement No 638809 Ð AIDA).  We thank Volker Springel for making his \GADGETtwo\ and \Subfind\ codes available to us and Peter Behroozi for making his \ROCKSTAR\ code available.  We also thank N. Gnedin for several very useful comments on our manuscript.

%% file: appendix.tex
\begin{table*}
\begin{minipage}{170mm}
\begin{center}
\begin{tabular}{|ccccccc}
\hline
Parameter  & Watson Fit & \Tiamat\ Fit & covariance with $A$  & covariance with $\alpha$ & covariance with $\beta$ & covariance with $\gamma$\\
\hline
$A$			& 0.282 & 0.0333	& $6.89{\times}10^{-5}$ & $1.65{\times}10^{-4}$ & $-1.93{\times}10^{-2}$ & $1.55{\times}10^{-5}$ \\
$\alpha$		& 2.163 & 1.153	& $1.65{\times}10^{-4}$ & $5.10{\times}10^{-4}$ & -$4.77{\times}10^{-2}$ & $6.82{\times}10^{-5}$ \\
$\beta$		& 1.406 & 12.33	& $-1.93{\times}10^{-2}$ & $-4.77{\times}10^{-2}$ & $5.80$ & $-4.81{\times}10^{-3}$ \\ 
$\gamma$	& 1.210 & 1.01		& $1.55{\times}10^{-5}$ & $6.82{\times}10^{-5}$ & $-4.81{\times}10^{-3}$ & $1.29{\times}10^{-5}$ \\ 
\hline
\end{tabular}
\caption{Best fitting values for and covariance between the \citet{Watson:2013p2555} mass function parameters as fit to \Subfind\ halos extracted from the \Tiamat\ suite of Planck-2015 EoR simulations. \label{table-mass_function_fit}}
\end{center}
\end{minipage}
\end{table*}

In this section we present additional details regarding the results of the friends-of-friends (FoF) halo mass function fitting that we present and discuss in Section \ref{sec-mass_sub_functions}.  We seek to test the parameterised universal FoF halo mass function presented by \citet{Watson:2013p2555} with the \Tiamat\ simulation suite.  This mass function follows the convention introduced by \citet[][see \citealt{Lukic:2007} for a good review]{Jenkins:2001p2646} whereby $n(M,z)$, the number density of FoF halos with mass $M$ at redshift $z$, is separated into a `scaled-mass function' component $f(\sigma,z)$ expected to be independent with redshift \citep[as predicted by the analytic theory of][and its extensions]{Press:1974p2491} and terms which encapsulate the linear growth of the matter density field

\begin{equation}\label{eqn-mass_function_definition}
\frac{dn}{d\log M}{=}\frac{\rho_b}{M} f(\sigma,z) \frac{d\ln\sigma^{-1}}{d\log M}
\end{equation}

\noindent where $\rho_b(z)$ is the mean background matter density as a function of redshift and $\sigma$ is the variance of the linearly evolved matter density field smoothed with a spherical tophat on a scale $R$ encompassing (on average) the halo mass $M$ (\ie\ $R{=}3M/[4\pi\rho_b(z)]^{1/3}$).  

As simulations have become increasingly more precise and halo finding approaches increasingly diverse, a large number of parameterisations for $f(\sigma,z)$ have been proposed with varying degrees of complexity and redshift evolution.  A `universal' (\ie\ redshift-independent) form for the mass function of structures extracted with the FoF algorithm was presented by \citet{Watson:2013p2555} with the form

\begin{equation}\label{eqn-Watson}
f(\sigma){=}A\left(\frac{\beta}{\sigma}+1\right)^\alpha e^{-\gamma/\sigma^2}
\end{equation}

\noindent where $A$ and $\beta$ are effectively two `normalisation parameters' and $\alpha$ and $\gamma$ are effectively two `shape parameters'.  We have fit this functional form to the mass functions presented in Figure \ref{fig-mass_functions} extracted from \Tiamat,\TinyTiamat\ and \MediTiamat\ at redshifts $z{=}5,7.5,10,15$ and $24$.  To do so we have used the implementation of the Metropolis-Hastings Monte-Carlo Markov Chain algorithm first presented in \citet{Poole:2013p1849} and utilised in several studies since.  We have sought to minimise systematic differences in our comparison to the \citet{Watson:2013p2555} fitting results.  We have thus followed their approach and restricted our fit to halos with more than 1000 particles.  We have also corrected for finite box size effects using the method employed by \citet{Lukic:2007} and \citet{Bhattacharya:2011p2652} (whereby fluctuations on scales larger than the box size are excluded from the tophat filtering calculation which maps halo mass to the variance of the matter density field), used the expression for 1$\sigma$ Poisson uncertainties introduced by \citet{Heinrich:2003}
\begin{equation}\label{eqn-Poisson_1sigma}
\sigma_{\pm}{=}\sqrt{N+\frac{1}{4}}{\pm}\frac{1}{2}
\end{equation}

\noindent for computing the $\chi^2$ likelihoods that we use for the fitting and applied a slight resolution correction to our FoF masses of the form
\begin{equation}\label{eqn-M_correction}
M_{\rm FoF}{=}m_p N_p \left(1-N_p^{-0.6}\right)
\end{equation}

\noindent where $N_p$ is the number of particles in the FoF halo and $m_p$ is the particle mass.

\begin{figure*}
\begin{center}
\includegraphics[width=170mm]{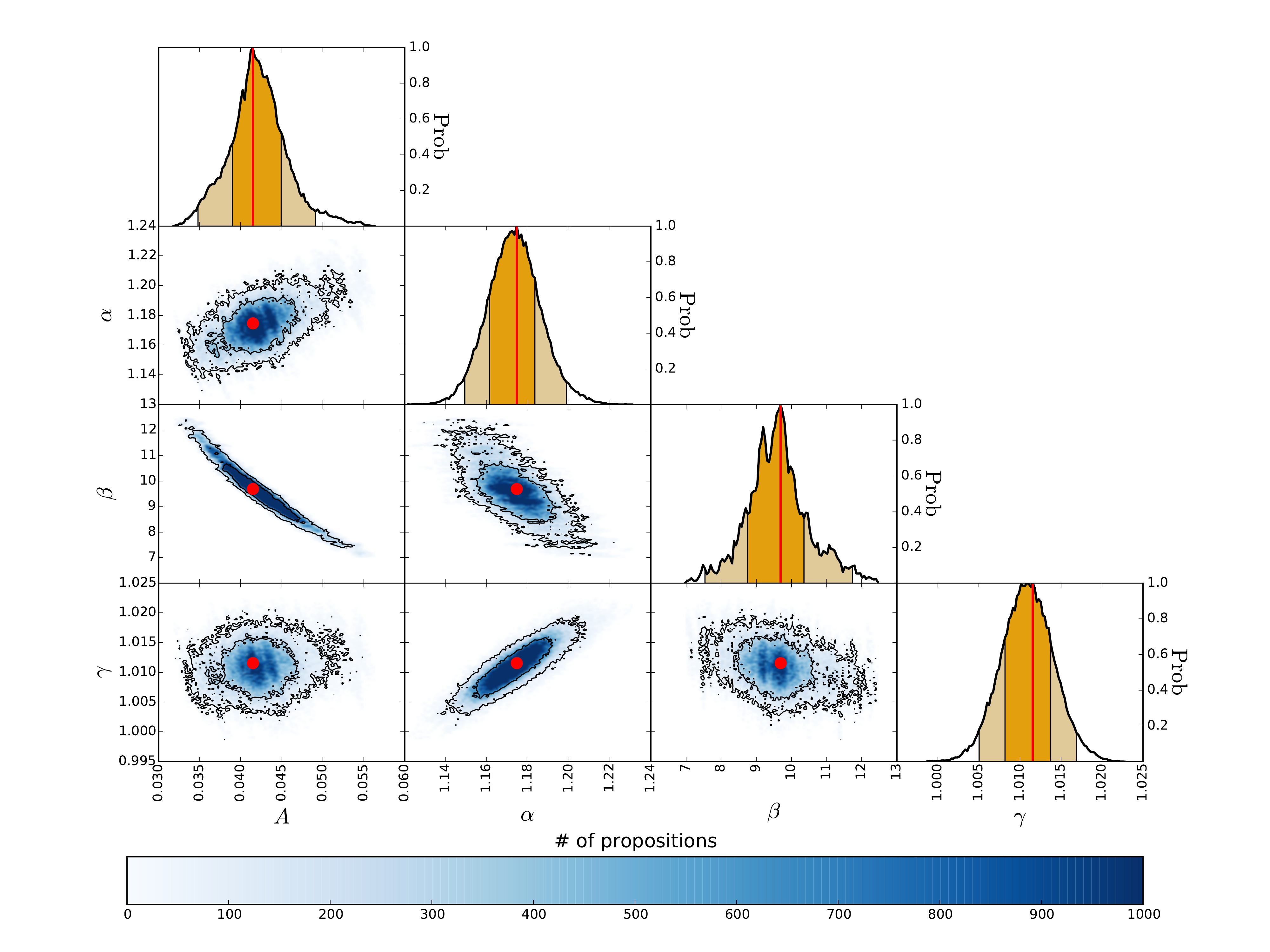}
\caption{One and two-dimensional projections of the posterior distribution function (pdf) of our MCMC fit of the parameterised universal FoF halo mass function presented by \citet[][see Equation \ref{eqn-Watson}]{Watson:2013p2555} to the FoF halo mass functions extracted from the \Tiamat\ simulation suite.  Blue-scale images show the pdf in terms of the number of propositions used to sample it, with black contours showing 68\% and 95\% confidence regions.  Dark and light orange shaded regions show the 68\% and 95\% confidence intervals of the one-dimensional projections respectively.  Red points and lines mark the most probable location in this parameter space, which we quote as the best fit parameters of our fit. \label{fig-pdf_matrix}}
\end{center}
\end{figure*}

\begin{figure}
\includegraphics[width=80mm]{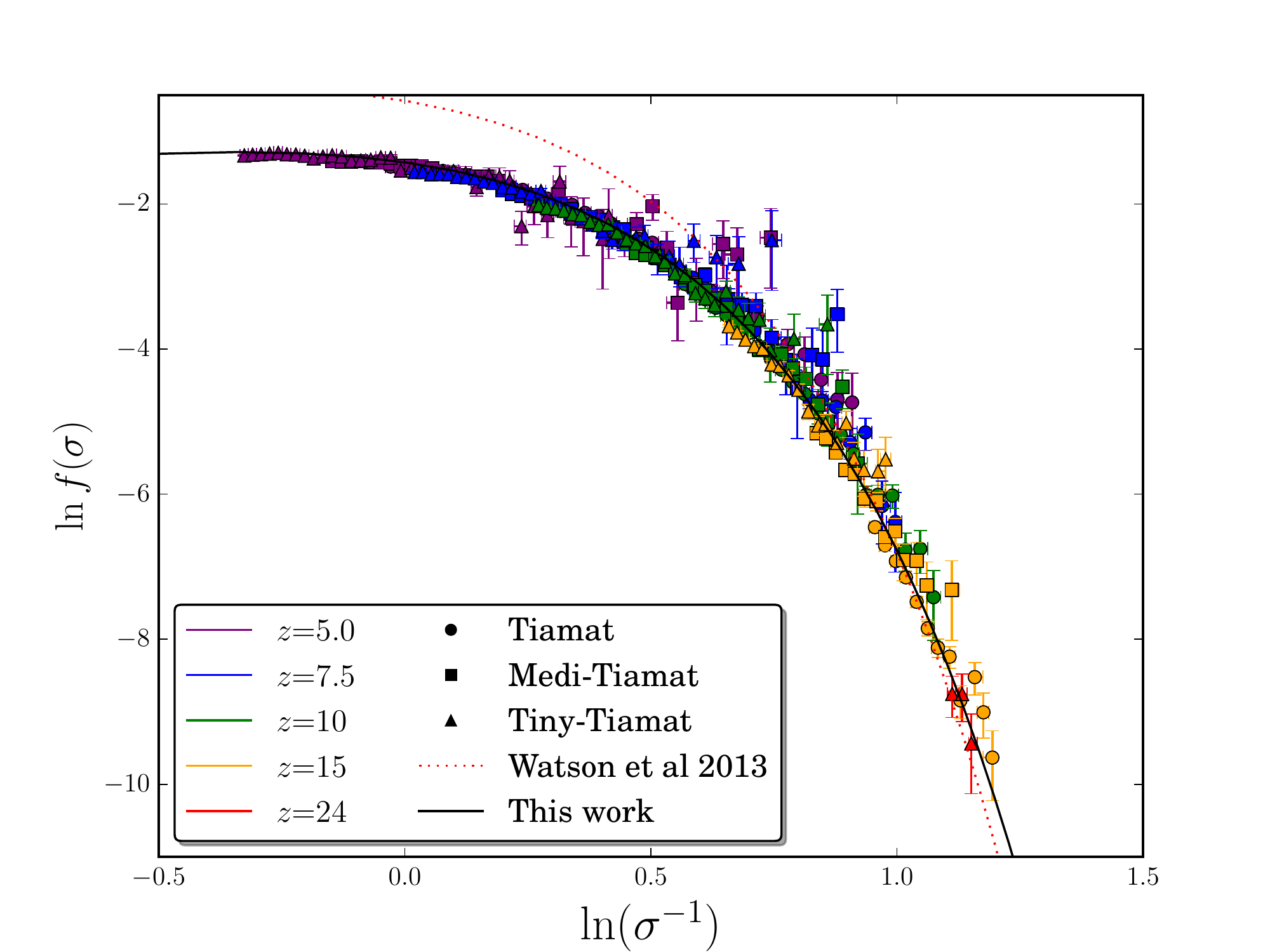}
\caption{Comparison of the scaled mass function of \citet{Watson:2013p2555} to data from \Tiamat\ (shown by the coloured data points, following the same format as Figure \ref{fig-mass_functions}) and to the parameterisation derived from \Tiamat\ and presented in this work. \label{fig-f_sigma_comparison}}
\end{figure}

We present the posterior distribution function (PDF) which emerges from this fit in Figure \ref{fig-pdf_matrix} and compare the parametrisation of Equation \ref{eqn-Watson}, as obtained across all redshifts by \citet{Watson:2013p2555}, to that obtained here for $z{\ge}5$.  A table of fitted values and their covariance are presented in Table \ref{table-mass_function_fit}.

At the low-mass (high-$\sigma$) end of the fit shown in Figure \ref{fig-f_sigma_comparison}, we find some significant differences between the \Tiamat\ results and the predictions of \citet{Watson:2013p2555} but good agreement otherwise.  Why then are our best-fit parameters so different?  We can see from Figure \ref{fig-pdf_matrix} that very strong degeneracies exist between the normalisation parameters ($A$ and $\beta$) and shape parameters ($\alpha$ and $\gamma$) of Equation \ref{eqn-Watson}.  Taking the shape parameters first, the suppression of low-mass systems requires a reduction of $\alpha$ to flatten the slope and an adjustment to $A$ to alter the low-mass normalisation.  In the case of the normalisation parameters, their product directly sets the normalisation of $f(\sigma)$ at the low-$\sigma$ end.  The product of these two parameters is strongly constrained by the data, as illustrated by the $A\beta{\sim}$constant form of the PDF in Figure \ref{fig-pdf_matrix}.  Given the adjustments discussed above which are needed to fit the high-$\sigma$ end of the function, our fit becomes pushed to a different part of this degeneracy.  While our best fit parameters for $A$ and $\beta$ are very different from the \citet{Watson:2013p2555} values, the products are very similar with our fit yielding $A\beta{=}0.407$ and the \citet{Watson:2013p2555} values yielding $A\beta{=}0.396$.  This reflects very similar results between the two studies at the high-mass (or low-$\sigma$) end, despite these very different fit values.